\documentclass[12pt]{article}
\usepackage{epsfig}
\usepackage{graphicx}
\usepackage{amsmath}

\begin{document}

 \newcommand{\beq}{\begin{equation}}
\newcommand{\eeq}{\end{equation}}
\newcommand{\bea}{\begin{eqnarray}}
\newcommand{\eea}{\end{eqnarray}}
\newcommand{\beqn}{\begin{eqnarray}}
\newcommand{\eeqn}{\end{eqnarray}}
\newcommand{\beas}{\begin{eqnarray*}}
\newcommand{\eeas}{\end{eqnarray*}}
\newcommand{\defi}{\stackrel{\rm def}{=}}
\newcommand{\non}{\nonumber}
\newcommand{\bquo}{\begin{quote}}
\newcommand{\enqu}{\end{quote}}
\newcommand{\qt}{\tilde q}
\newcommand{\m}{\tilde m}
\newcommand{\trho}{\tilde{\rho}}
\newcommand{\tn}{\tilde{n}}
\newcommand{\tN}{\tilde N}

\newcommand{\gsim}{\lower.7ex\hbox{$
\;\stackrel{\textstyle>}{\sim}\;$}}
\newcommand{\lsim}{\lower.7ex\hbox{$\;\stackrel{\textstyle<}{\sim}\;$}}


\def\de{\partial}
\def\Tr{ \hbox{\rm Tr}}
\def\const{\hbox {\rm const.}}
\def\o{\over}
\def\im{\hbox{\rm Im}}
\def\re{\hbox{\rm Re}}
\def\bra{\langle}\def\ket{\rangle}
\def\Arg{\hbox {\rm Arg}}
\def\Re{\hbox {\rm Re}}
\def\Im{\hbox {\rm Im}}
\def\diag{\hbox{\rm diag}}


\def\QATOPD#1#2#3#4{{#3 \atopwithdelims#1#2 #4}}
\def\stackunder#1#2{\mathrel{\mathop{#2}\limits_{#1}}}
\def\stackreb#1#2{\mathrel{\mathop{#2}\limits_{#1}}}
\def\Tr{{\rm Tr}}
\def\res{{\rm res}}
\def\Bf#1{\mbox{\boldmath $#1$}}
\def\balpha{{\Bf\alpha}}
\def\bbeta{{\Bf\beta}}
\def\bgamma{{\Bf\gamma}}
\def\bnu{{\Bf\nu}}
\def\bmu{{\Bf\mu}}
\def\bphi{{\Bf\phi}}
\def\bPhi{{\Bf\Phi}}
\def\bomega{{\Bf\omega}}
\def\blambda{{\Bf\lambda}}
\def\brho{{\Bf\rho}}
\def\bsigma{{\bfit\sigma}}
\def\bxi{{\Bf\xi}}
\def\bbeta{{\Bf\eta}}
\def\d{\partial}
\def\der#1#2{\frac{\d{#1}}{\d{#2}}}
\def\Im{{\rm Im}}
\def\Re{{\rm Re}}
\def\rank{{\rm rank}}
\def\diag{{\rm diag}}
\def\2{{1\over 2}}
\def\ntwo{${\mathcal N}=2\;$}
\def\nfour{${\mathcal N}=4\;$}
\def\none{${\mathcal N}=1\;$}
\def\ntwot{${\mathcal N}=(2,2)\;$}
\def\ntwoo{${\mathcal N}=(0,2)\;$}
\def\x{\stackrel{\otimes}{,}}

\def\ba{\beq\new\begin{array}{c}}
\def\ea{\end{array}\eeq}
\def\be{\ba}
\def\ee{\ea}
\def\stackreb#1#2{\mathrel{\mathop{#2}\limits_{#1}}}

\def\Tr{{\rm Tr}}
\newcommand{\cpn}{CP$(N-1)\;$}
\newcommand{\wcpn}{wCP$_{N,\tilde{N}}(N_f-1)\;$}
\newcommand{\wcpd}{wCP$_{\tilde{N},N}(N_f-1)\;$}
\newcommand{\vp}{\varphi}
\newcommand{\pt}{\partial}
\newcommand{\ve}{\varepsilon}
\renewcommand{\theequation}{\thesection.\arabic{equation}}

\setcounter{footnote}0

\vfill

\begin{titlepage}

\begin{flushright}
FTPI-MINN-10/01, UMN-TH-2831/10\\
February 1, 2010
\end{flushright}

\begin{center}
{  \Large \bf  
Non-Abelian Confinement  in \boldmath{\ntwo} Supersymmetric QCD:
\\[1.5mm]
Duality and Kinks on  Confining Strings
  }

\vspace{4mm}

 {\large
 \bf    M.~Shifman$^{\,a}$ and \bf A.~Yung$^{\,\,a,b}$}
\end {center}

\begin{center}


$^a${\it  William I. Fine Theoretical Physics Institute,
University of Minnesota,
Minneapolis, MN 55455, USA}\\
$^{b}${\it Petersburg Nuclear Physics Institute, Gatchina, St. Petersburg
188300, Russia
}
\end{center}

\begin{center}
{\large\bf Abstract}
\end{center}

Recently we observed a crossover transition (in the Fayet--Iliopoulos parameter)
from weak to strong coupling in \ntwo supersymmetric QCD with the U($N$) gauge group
and  $N_f>N $ quark flavors. At strong coupling this theory can be described by 
a dual non-Abelian weakly coupled  SQCD with 
the dual gauge group  U$(N_f-N)$ and  $N_f$ light dyon flavors.
Both theories support non-Abelian strings.
We continue the study of confinement dynamics in these theories,
in particular, metamorphoses of excitation spectra,
from a different side. A number of results obtained previously
are explained, enhanced and supplemented by analyzing 
the world-sheet dynamics on the non-Abelian confining strings. The world-sheet theory is
the two-dimensional \ntwot supersymmetric
weighted CP$(N_f-1)$  model. 
We explore the vacuum structure and kinks on the world sheet, corresponding to confined monopoles
in the bulk theory. We show that (in the equal quark mass limit) these kinks fall into
the fundamental representation of the unbroken global SU$(N)\times$SU$(N_f-N)\times$U(1) group.
This result confirms the presence of ``extra" stringy meson states in the adjoint
representation of the global group in the bulk theory. The non-Abelian bulk duality is in one-to-one correspondence
with a duality taking place in the \ntwot supersymmetric
weighted CP$(N_f-1)$  model.

\vspace{2cm}

\end{titlepage}

 \newpage

\tableofcontents

\newpage

\section {Introduction}
\label{intro}
\setcounter{equation}{0}

The standard scenario for color confinement suggested in the 1970s by  Nambu, Mandelstam and 't Hooft  \cite{mandelstam}
is based on the dual Meissner effect. In this scenario, 
upon condensation of monopoles,  chromoelectric flux tubes (strings)  of the Abrikosov--Nielsen--Olesen
(ANO) type  \cite{ANO} are formed. This must lead to confinement of quarks  attached to
the endpoints of confining strings.

Much later it was shown by Seiberg and Witten \cite{SW1,SW2} that this scenario is 
indeed realized in \ntwo 
supersymmetric QCD in the monopole vacua. A more careful examination shows, however,
that this confinement is Abelian\,\footnote{By  non-Abelian
confinement we mean such dynamical regime in which at distances
of the flux tube formation all gauge bosons are equally important,
while Abelian confinement occurs when the relevant gauge dynamics at such distances is Abelian.
Note that Abelian confinement can take place in non-Abelian theories.
The Seiberg--Witten solution is just one example. Another example is the Polyakov three-dimensional confinement
\cite{Polyakov} in the Georgi--Glashow model.}
\cite{DS,HSZ,Strassler,VY,Yrev}. The reason is that the non-Abelian gauge group of
underlying \ntwo SQCD (say,  SU$(N)$) is broken down to Abelian U(1)$^{N-1}$ subgroup by
condensation of adjoint scalars in the strongly coupled monopole vacua.
Further condensation of monopoles occurs essentially in the U(1) theory.

A non-Abelian mechanism for  confinement in four dimensions was recently proposed in \cite{SYdual}. 
In this paper   we considered \ntwo SQCD with the U($N$) gauge group and
$N_f$ flavors of fundamental quark hypermultiplets, $N<N_f<2N$. This theory is endowed with the 
Fayet--Iliopoulos (FI) \cite{FI} term $\xi$ which singles out a vacuum in which  $r=N$ scalar quarks
condense.  At large $\xi$ this theory is at weak coupling.
In the limit of equal quark masses it supports non-Abelian strings
 \cite{HT1,ABEKY,SYmon,HT2} (see also the review papers \cite{Trev,Jrev,SYrev,Trev2}). 
Formation of these strings  leads to confinement of monopoles. 
In fact, in the Higgsed U$(N)$ gauge theories the monopoles
become junctions of two distinct elementary non-Abelian strings.

In \cite{SYdual} (see
also \cite{SYcross,SYcrossp}) 
we demonstrated
that upon reducing the FI parameter $\xi$ the theory
goes through a crossover transition into a strongly coupled phase which can be described
in the infrared 
in terms of weakly coupled {\em dual} \ntwo SQCD with the U($\tN$) gauge group and $N_f$ flavors of
light dyons,\footnote{This is in a perfect agreement with the
results obtained in \cite{APS} where the SU$(\tN)$ dual  gauge group  was identified
at the root of a baryonic Higgs branch in the   SU($N$) gauge theory with massless (s)quarks.}
where 
\beq
\tN=N_f-N.
\label{tN}
\eeq
This non-Abelian \ntwo duality is conceptually similar to Seiberg's duality in \none SQCD
\cite{Sdual,IS}
where the emergence of the dual SU$(\tN)$ group was first observed.

The dual theory also supports non-Abelian strings formed due to condensation of light dyons.
Moreover, these  latter strings still confine monopoles, 
rather than quarks \cite{SYdual}. Thus, 
the \ntwo non-Abelian duality is {\em not}  the electromagnetic duality. 
It is the monopoles that
are confined  both, in the original and dual theories. The reason for this is as follows. Light
dyons which condense in the dual theory (in addition to magnetic charges) 
carry weight-like electric charges,
(i.e. the quark charges). Therefore, the strings formed through the condensation of these dyons
can confine only the states with the root-like magnetic charges, i.e. 
the monopoles, see  \cite{SYdual} for more details.

\begin{figure}
\epsfxsize=6cm
\centerline{\epsfbox{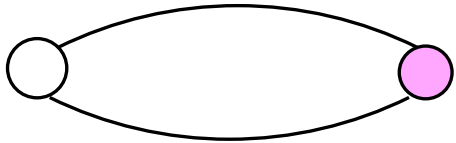}}
\caption{\small Meson formed by monopole-antimonopole pair connected by two strings.
Open and closed circles denote the monopole and antimonopole, respectively.}
\label{figmeson}
\end{figure}

Our mechanism of non-Abelian confinement works as follows. There is no confinement of
color-electric charges, to begin with. The color-electric charges of quarks (or gauge bosons)
are Higgs-screened. In the domain of small $\xi$ (where the dual description
is applicable) the quarks and
gauge bosons of the original theory  decay into the monopole-antimonopole 
pairs at the curves of marginal stability (CMS).
At small but nonvanishing $\xi$ the (anti)monopoles forming the pair
 cannot abandon
each other  because they are confined. In other words, the original quarks and gauge bosons 
evolve in the 
strong coupling domain of small $\xi$  into ``stringy mesons" with two constituents 
being connected by two strings as shown in  Fig.~\ref{figmeson}, see \cite{SYrev} for a detailed discussion
of these stringy mesons. 
The color-magnetic charges are confined in the theory under consideration; the mesons they form are expected
to lie on Regge trajectories. 

One might think that this pattern of confinement has little to do with 
what we have in real life
 because the real-world mesons can be in the adjoint representation of the global flavor
group, while the
monopoles we discuss at first glance seem to be neutral under the flavor group.
If so, the monopole-antimonopole pairs
could  produce only flavor-singlet mesons. 
In this paper we address this problem and show that this naive guess is incorrect.
We demonstrate that deep in the non-Abelian quantum regime the confined monopoles  are in the
{\em fundamental representation} of the global  group. Therefore, the  monopole-antimonopole
mesons can be both, in the adjoint and singlet representation of the flavor group. 

If in Ref.~\cite{SYdual} we explored the non-Abelian duality and the evolution of
spectra in the $\xi$ transition from the standpoint of the four-dimensional
bulk theories, in this work we will invoke a totally different and seemingly quite powerful tool.
Our strategy is to expand and supplement the bulk theory analysis
\cite{SYdual} by studies in the world-sheet theories on the
non-Abelian strings supported by the bulk theory. As we know from our previous work, in the
BPS sector there should exist a one-to-one correspondence between
the bulk and world-sheet results. Therefore, metamorphoses of spectra can be studied
in the world sheet theory, providing us with additional information. In particular, the
latter should   undergo its own crossover transition into a dual world-sheet theory.
The obvious strategical advantage is a relative simplicity of two-dimensional theories compared
to their four-dimensional progenitor.

The main feature of the non-Abelian strings 
is the presence of orientational zero modes. Dynamics of these orientational zero modes 
(uplifted to two-dimensional fields)
 can be  described, at low energies,  by an effective two-dimensional  sigma model  on the 
 string world sheet. Particular details of this model depend on the bulk parameters.
For instance, in the simplest case of  \ntwo SQCD with the U($N$) gauge
 group and $N_f=N$ (s)quarks, one obtains 
the \ntwot supersymmetric \cpn
 model on the world sheet \cite{HT1,ABEKY,SYmon,HT2}. If $N_f>N$
 it is the weighted CP$(N_f-1)$  model that we get.

In our previous works we revealed a number of ``protected" quantities, such as the mass
of the (confined) monopoles. These
parameters are calculable both,
in the bulk theory and on the world sheet, with one and the same result.
The first example of this remarkable correspondence was the explanation of the
coincidence of the BPS spectrum of monopoles in four-dimensional theory in 
the $r=N$ vacuum 
on the Coulomb branch
at $\xi=0$ (given by the exact Seiberg--Witten solution \cite{SW2}) with the
BPS spectum of kinks in the \ntwot supersymmetric \cpn
 model. This coincidence was noted in \cite{Dorey,DoHoTo}.
 The explanation   
 \cite{SYmon,HT2} is : (i) the confined monopoles of the bulk theory (represented
 by two-string junctions) are seen as kinks interpolating between two different vacua in the
sigma model on the string world sheet;  (ii) the masses of the BPS monopoles cannot depend on
the nonholomorphic parameter $\xi$.

Later on various deformations of the bulk theory were considered and their responses  in the 
sigma model on the string world sheet were found, for reviews see~\cite{SYrev,Trev2}. In all cases
the bulk physics   is ``projected" onto the world sheet physics. The protected quantities come out the same.
However, technical/calculational  aspects are much simpler in the two-dimensional
world-sheet theory than in the four-dimensional bulk theory. Therefore, it is beneficial
to use the above correspondence not only in the direction from four to two dimensions
(the preferred direction in the past), but in the opposite direction too.
In the present paper we exploit this idea and
 study confined monopoles of the bulk theory
in terms of kinks of the effective theory on the string world sheet.
Two-dimensional CP models are well understood even at strong coupling. We
 use this knowledge to extract information on confined monopoles of the bulk theory. 

If $N_f>N$ the non-Abelian
strings are semilocal (see \cite{AchVas} for a review), and  the effective sigma model on their world sheet is the
\ntwot weighted CP$(N_f-1)$ model on a toric manifold \cite{HT1,SYsem,Jsem}.
The bulk duality observed in \cite{SYdual} must be
in one-to-one correspondence with
a two-dimensional duality in
the weighted CP model.

At large $\xi$, 
internal dynamics of the semilocal non-Abelian strings is described by
the sigma model of $N$ orientational and $\tN$ size moduli, while
at small $\xi$ the roles of orientational and size moduli interchange.
Two dual weighted CP models transform into each other upon changing the sign of the coupling
constant \cite{SYdual}.
The BPS kink spectra in these two dual sigma models (describing the confined 
monopoles of the bulk theory) coincide.

In this paper we  study the kinks in the weighted CP model  in detail, calculate their spectra and show that,
in the equal-mass limit of strong coupling (inside CMS), the kinks fall in fundamental
representation of the global  symmetry group (aka flavor group). 
This confirms our picture of confinement in the bulk theory.
In particular, the mesons shown in Fig.~\ref{figmeson} belong either to the adjoint or to singlet
representation of the flavor group, as was expected.

The fact  that  the kinks in the quantum limit form a fundamental representation
of the global group is not that surprising. Say, in the  \ntwot supersymmetric \cpn models it was known for a 
long time \cite{W79,HoVa}. Here we generalize this result to the case of the \ntwot weighted CP models and
translate it in terms of the confined monopoles of the bulk theory.

The paper is organized as follows. In Sec.~\ref{secbulk} we briefly review 
duality in the bulk  four-dimensional theory and discuss evolution of 
excitation spectra in passing from one side of duality to another,
through the crossover domain.
Section \ref{secWCP} is devoted to the world-sheet theory on the non-Abelian strings -- the
weighted CP model. We outline 
 its two versions related to each other by the world-sheet duality. In 
Sec.~\ref{secclasslim} we treat 
the  (semi)classical limit of the world-sheet theory, examining both components of the  dual pair.
 In
Sec.~\ref{secsup}
we consider the exact superpotential and the semiclassical BPS spectrum at large and intermediate
values of $|\Delta m_{A,B}|$, i.e.  outside CMS. In
Sec.~\ref{secmirror} 
we formulate and explore a mirror representation for both dual theories. In 
Sec.~\ref{seckink} 
we study kinks using the mirror
representation. We calculate their spectra  and count the
 number of kinks at strong coupling, inside CMS. In Sec.~\ref{secmoral} we 
 translate our two-dimensional results   in four-dimensional bulk theory, i.e. interpret them 
in terms of strings and confined monopoles of the bulk theory.
Section~\ref{concl}
summarizes our conclusions.

\section{Bulk duality}
\label{secbulk}
\setcounter{equation}{0}

This section presents a brief review of the bulk non-Abelian duality \cite{SYdual}
and introduces all relevant notation (which is also summarized in \cite{SYrev}).
The bulk theory is  \ntwo SQCD with the U($N$) gauge group and
$N_f$ flavors of fundamental quark hypermultiplets ($N<N_f<2N$). 

\subsection{Bulk theory at large $\xi$}

The field content is as follows. The \ntwo vector multiplet
consists of the  U(1)
gauge field $A_{\mu}$ and the SU$(N)$  gauge field $A^a_{\mu}$,
where $a=1,..., N^2-1$, and their Weyl fermion superpartners
 plus
complex scalar fields $a$, and $a^a$ and their Weyl superpartners.
The $N_f$ quark multiplets of  the U$(N)$ theory consist
of   the complex scalar fields
$q^{kA}$ and $\tilde{q}_{Ak}$ (squarks) and
their   fermion superpartners, all in the fundamental representation of 
the SU$(N)$ gauge group.
Here $k=1,..., N$ is the color index
while $A$ is the flavor index, $A=1,..., N_f$. We will treat $q^{kA}$  
as a rectangular matrix with $N$ rows and $N_f$ columns.

This theory is endowed with
the FI term $\xi$ which singles out the vacuum in which  $r=N$ squarks condense.
Consider, say, the (1,2,...,$N$) vacuum in which the first $N$ flavors develop 
vacuum expectation values
(VEVs),
\beqn
\langle q^{kA}\rangle &=&\sqrt{
\xi}\,
\left(
\begin{array}{cccccc}
1 & \ldots & 0 & 0 & \ldots & 0\\
\ldots & \ldots & \ldots  & \ldots & \ldots & \ldots\\
0 & \ldots & 1 & 0 & \ldots & 0\\
\end{array}
\right),
\qquad \langle \bar{\tilde{q}}^{kA}\rangle =0,
\nonumber\\[4mm]
k&=&1,..., N\,,\qquad A=1,...,N_f\, .
\label{qvev}
\eeqn
In this vacuum the
adjoint fields also develop  
VEVs, namely,
\beq
\left\langle \left(\frac12\, a + T^a\, a^a\right)\right\rangle = - \frac1{\sqrt{2}}
 \left(
\begin{array}{ccc}
m_1 & \ldots & 0 \\
\ldots & \ldots & \ldots\\
0 & \ldots & m_N\\
\end{array}
\right),
\label{avev}
\eeq
where $m_A$ are quark mass parameters.

For generic values of $m_A$'s, the VEVs (\ref{avev}) break the  SU$(N)$ subgroup of the gauge 
group down to U(1)$^{N-1}$. However, in the special limit
\beq
m_1=m_2=...=m_{N_f},
\label{equalmasses}
\eeq
the  SU$(N)\times$U(1) gauge group remains  unbroken by the adjoint field.
In this limit the theory acquires a global flavor SU$(N_f)$ symmetry.

While the adjoint VEVs do not break the SU$(N)\times$U(1) gauge group in the limit
(\ref{equalmasses}), 
the quark condensate  (\ref{qvev}) results in  the spontaneous
breaking of both gauge and flavor symmetries.
A diagonal global SU$(N)$  combining the gauge SU$(N)$ and an
SU$(N)$ subgroup (which rotates first $N$ quarks) of the flavor SU$(N_f)$
group survives, however. Below we will refer to this diagonal
global symmetry as to $ {\rm SU}(N)_{C+F}$.
More exactly, the pattern of breaking of the
color and flavor symmetry 
is as follows: 
\beq
{\rm U}(N)_{\rm gauge}\times {\rm SU}(N_f)_{\rm flavor}\to  
{\rm SU}(N)_{C+F}\times  {\rm SU}(\tilde{N})_F\times {\rm U}(1)\,,
\label{c+f}
\eeq
where $\tilde{N}=N_f-N$.
The phenomenon of color-flavor locking takes place in the vacuum. 
The global SU$(N)_{C+F}$ group is responsible  for
formation of the non-Abelian strings (see below).
For unequal quark masses in (\ref{avev}) the  global symmetry  (\ref{c+f}) is broken down to 
U(1)$^{N_f-1}$.

Since the global (flavor) SU$(N_f)$ group is broken by the quark VEVs anyway we can consider
the following mass splitting:
\beq
m_P=m_{P'}, \qquad m_K=m_{K'}, \qquad m_P-m_K=\Delta m
\label{masssplit}
\eeq
where $P,P'=1, ..., N$ and $K,K'=N+1, ..., N_f$.\footnote{A generic mass difference $m_A-m_B$ 
(for all $A,B = 1,2, ..., N_f$)
will be referred to as $\Delta m_{AB}$ below,
while $\Delta m$ is reserved for $m_P-m_K$, ($P=1,2, ..., N$, $K=N+1, ..., N_f$).
  In Ref.~\cite{SYcrossp}
the mass differences inside the first group
(or inside the second group) were called $\Delta M_{\rm inside}$.
The mass differences $m_P-m_K$ were referred to as 
$\Delta M_{\rm outside}$.} 
This mass splitting respects the global
group (\ref{c+f}) in the $(1,2,...,N)$ vacuum. This vacuum then becomes  isolated.
No Higgs branch  develops.  We will often use this limit below.

Now let us discuss the mass spectrum in our theory. Since
both U(1) and SU($N$) gauge groups are broken by squark condensation, all
gauge bosons become massive. In fact, at nonvanishing $\xi$, both the quarks and adjoint scalars  
combine  with the gauge bosons to form long \ntwo supermultiplets \cite{VY},  for a review see
\cite{SYrev}.
Note that all states come in representations of the unbroken global
 group (\ref{c+f}), namely, the singlet and adjoint representations
of SU$(N)_{C+F}$
\beq
(1,\, 1), \quad (N^2-1,\, 1),
\label{onep}
\eeq
 and in the bifundamental representations
\beq
 \quad (\bar{N},\, \tN), \quad
(N,\, \bar{\tN})\,,
\label{twop}
\eeq
 where in (\ref{onep}) and (\ref{twop}) we mark representation with respect to two 
non-Abelian factors in (\ref{c+f}). The singlet and adjoint fields are the gauge bosons,
and the first $N$ flavors of the squarks $q^{kP}$ ($P=1,...,N$), together with their fermion superpartners.
The bifundamental fields are the quarks $q^{kK}$ with $K=N+1,...,N_f$.
These quarks transform in the two-index representations of the global
group (\ref{c+f}) due to the color-flavor locking.

At large $\xi$ this theory is at weak coupling. Namely, the condition
\beq
\xi\gg \Lambda\, ,
\label{weakcoupling}
\eeq
ensures weak coupling in the SU$(N)$ sector because
the SU$(N)$ gauge coupling does not run below the scale of the quark VEVs
which is determined by $\sqrt\xi$. Here $\Lambda$ is the dynamical scale of the SU($N$) gauge theory. More explicitly,
\beq
\frac{8\pi^2}{g^2_2 (\xi)} =
(N-\tN )\ln{\frac{g_2\sqrt{\xi}}{\Lambda}}\gg 1 \,,       \rule{8mm}{0mm}
\label{4coupling}
\eeq
where $g_2^2$ is the coupling constant of the SU$(N)$ sector.

\subsection{Duality}
\label{secbulkduality}

As was shown in \cite{SYdual}, at  $\sqrt\xi \sim \Lambda$ the theory goes through a crossover transition
to the strong coupling regime. At small  $\xi$ ($\sqrt\xi \ll \Lambda$) this regime can be described
in terms of weakly coupled dual \ntwo SQCD, with the gauge group
\beq
{\rm U}(\tN)\times {\rm U}(1)^{N-\tN}\,,
\label{dualgaugegroup}
\eeq
 and $N_f$ flavors of
light {\em dyons}. 
This non-Abelian \ntwo duality is similar to Seiberg's duality in \none supersymmetric QCD
\cite{Sdual,IS}. Later a dual non-Abelian gauge group SU$(\tN)$ was identified on the Coulomb branch 
at the root of a baryonic Higgs branch in the \ntwo supersymmetric  SU($N$) gauge theory with massless quarks
\cite{APS}.

Light dyons are in the fundamental representation of the gauge group
U$(\tN)$ and are charged under Abelian factors in (\ref{dualgaugegroup}). In addition, there are   
light dyons $D^l$ ($l=\tN+1, ..., N$) neutral under 
the U$(\tN)$ group, but charged under the
U(1) factors. A small but nonvanishing $\xi$ triggers condensation of all these
dyons,
\beqn
\langle D^{lA}\rangle \! \! &=&\!\!\sqrt{
\xi}\,
\left(
\begin{array}{cccccc}
0 & \ldots & 0 & 1 & \ldots & 0\\
\ldots & \ldots & \ldots  & \ldots & \ldots & \ldots\\
0 & \ldots & 0 & 0 & \ldots & 1\\
\end{array}
\right),
\quad \langle \bar{\tilde{D}}^{lA}\rangle =0,\quad l=1,...,\tN,
\nonumber\\[4mm]
\langle D^{l}\rangle &=& \sqrt{\xi}, \qquad \langle\bar{\tilde{D}}^{l}\rangle =0\,,
\qquad l=\tN +1, ..., N\,.
\label{Dvev}
\eeqn

Now, consider either  equal quark masses or the mass choice (\ref{masssplit}).
Both, the gauge  and flavor SU($N_f$) groups, are
broken in the vacuum. However, the color-flavor locked form of (\ref{Dvev}) guarantees that the diagonal
global SU($\tN)_{C+F}$ survives. More exactly, the  unbroken global group of the dual
theory is 
\beq
 {\rm SU}(N)_F\times  {\rm SU}(\tN)_{C+F}\times {\rm U}(1)\,.
\label{c+fd}
\eeq
Here SU$(\tN)_{C+F}$ is a global unbroken color-flavor rotation, which involves the
last $\tN$ flavors, while SU$(N)_F$ factor stands for the flavor rotation of the 
first $N$ dyons.
Thus, a color-flavor locking takes place in the dual theory too. Much in the same way as 
in the original theory, the presence of the global SU$(\tN)_{C+F}$ group 
is the  reason behind formation of the non-Abelian strings.
 For generic quark masses the  global symmetry  (\ref{c+f}) is broken down to 
U(1)$^{N_f-1}$. 

In the equal mass limit or for the mass choice (\ref{masssplit})
the global unbroken symmetry (\ref{c+fd}) of the dual theory at small
$\xi$ coincides with the global group (\ref{c+f}) present in the
$r=N$ vacuum of the original theory at large
$\xi$.  Note
however, that this global symmetry is realized in two distinct ways in two dual theories.
As was already mentioned, the quarks and U($N$) gauge bosons of the original theory at large $\xi$
come in the $(1,1)$, $(N^2-1,1)$, $(\bar{N},\tN)$, and $(N,\bar{\tN})$
representations of the global group (\ref{c+f}), while the dyons and U($\tN$) gauge 
bosons form $(1,1)$, $(1,\tN^2-1)$, $(N,\bar{\tN})$, and 
$(\bar{N},\tN)$ representations of (\ref{c+fd}). We see that the
adjoint representations of the $(C+F)$
subgroup are different in two theories. A similar phenomenon was detected in \cite{SYcross}
for the Abelian dual theory (i.e. $\tN=0$).
     
\begin{figure}
\epsfxsize=7cm
\centerline{\epsfbox{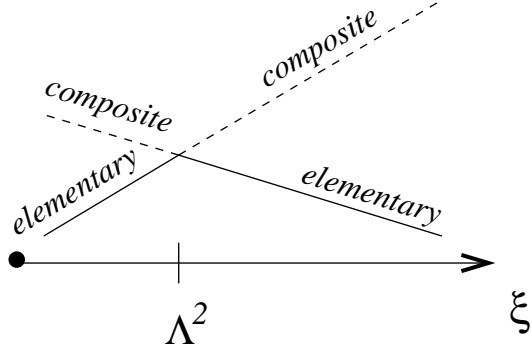}}
\caption{\small Evolution of the SU$(N)$ and SU$(\tN)$ gauge bosons and light quarks (dyons) vs. $\xi$. 
On both sides of the level crossing at $\xi=\Lambda^2$ the global groups are SU$(N)\times$SU$(\tN)$,
however, above $\Lambda^2$ it is SU$(N)_{C+F}\times$SU$(\tN)_F$ while
below $\Lambda^2$ it is SU$(N)_F\times$SU$(\tN)_{C+F}$.}
\label{figevol}
\end{figure}

This means that 
the quarks and gauge bosons
which form the  adjoint $(N^2-1)$ representation  
of SU($N$) at large $\xi$ and the dyons and gauge bosons which form the  adjoint $(\tN^2-1)$ representation  of SU($\tN$) at small $\xi$ are, in fact, {\em distinct} states.
The $(N^2-1)$  adjoints of SU($N$) become heavy 
and decouple as we pass from large to small $\xi$ 
 along the line (\ref{masssplit}). Moreover, some 
composite $(\tN^2-1)$ adjoints  of SU($\tN$), which are 
heavy  and invisible in the low-energy description at large $\xi$ become light 
at small $\xi$ and form the $D^{lK}$ dyons
 ($K=N+1,...,N_f$) and gauge bosons of U$(\tN)$. The phenomenon of level crossing
 takes place (Fig.~\ref{figevol}). Although this crossover is smooth in the full theory,
from the standpoint of the low-energy description the passage from  large to small $\xi$  means a dramatic change: the low-energy theories in these domains are 
completely
different; in particular, the degrees of freedom in these theories are different.

This logic leads us to the following conclusion. In addition to light dyons and gauge bosons 
 included in  the low-energy theory   at small $\xi$ we have
heavy  fields  which form the adjoint representation
$(N^2-1,1)$ of the global symmetry (\ref{c+fd}). These are screened  quarks 
and gauge bosons from the large $\xi$ domain.
 Let us denote them as $M_P^{P'}$ ($P,P'=1,...,N$). 

As was already explained in Sec.~\ref{intro},
at small $\xi$ they decay into the monopole-antimonopole 
pairs on the curves of marginal stability (CMS). \footnote{Strictly speaking,
such pairs 
can be  formed by monopole-antidyons and
 dyon-antidyons as well,
the dyons carrying root-like electric charges. In this paper we will call all these states
``monopoles". This is to avoid confusion with dyons which appear in Eq.~(\ref{Dvev}). The
latter dyons carry weight-like electric charges and, roughly speaking, behave as
quarks, see \cite{SYdual} for further details.}
This is in accordance with results obtained 
for \ntwo SU(2) gauge theories \cite{SW1,SW2,BF} on the Coulomb branch at zero $\xi$
(we confirm this result for the theory at hand in Sec.\ref{secmoral}).
The general rule is that the only states which exist at strong coupling inside CMS are those which can become massless on the Coulomb branch
\cite{SW1,SW2,BF}. For our theory these are light dyons shown in Eq.~(\ref{Dvev}),
gauge bosons of the dual U$(\tN)$ theory and monopoles.
 
 At small nonvanishing $\xi$ the
monopoles and antimonopoles produced in the decay process of adjoints $(N^2-1,1)$
 cannot escape from
each other and fly off to separate  because they are confined. Therefore, the quarks or gauge bosons in the 
strong coupling domain of small $\xi$ evolve into stringy mesons $M_P^{P'}$ ($P,P'=1, ..., N$) 
-- the  monopole-antimonopole
pairs connected  by two strings \cite{SYdual}  as shown in  Fig.~\ref{figmeson}. 

By the same token,  at large $\xi$, in addition to the light quarks and gauge bosons,
we  have heavy fields $M_K^{K'}$ ($K,K'=N+1, ..., N_f$), which  form the  adjoint $(\tN^2-1)$ representation  of SU($\tN$).
This is schematically depicted in Fig.~\ref{figevol}. 

The $M_K^{K'}$ states are (screened)  light
dyons and gauge bosons of the dual theory. At large $\xi$ they decay into
 monopole-antimonopole 
pairs and form stringy mesons  \cite{SYdual} shown in Fig.~\ref{figmeson}.

In \cite{SYdual}  we also conjectured that the fields $M_P^{P'}$ and $M_K^{K'}$ are Seiberg's meson fields
\cite{Sdual,IS},
which occur in the dual theory upon breaking of \ntwo supersymmetry by the mass-term
 superpotential $\mu[A^2 +(A^a)^2]$ for the adjoint fields in the limit 
$\mu\to\infty$. In this limit  our theory becomes \none SQCD.

We see that the picture of the non-Abelian confinement obtained  in \cite{SYdual} is based on the presence of 
extra stringy meson states -- the
monopole-antimonopole pairs -- bound by confining strings both in the weak and strong coupling
domains of  the bulk theory. These meson states fill representations  $(N^2-1,1)$
and $(1,\tN^2-1)$ of the global unbroken group at   small and large $\xi$,
respectively. Below  we confirm the presence of these stringy mesons by
studying the global quantum numbers of confined monopoles in both domains. To this end
we explore kinks in the \ntwot supersymmetric  weighted CP model on the   string world sheet.

We remind that the confined  monopoles of the bulk theory are presented by the junctions of
two elementary non-Abelian strings \cite{T,SYmon,HT2}. These elementary 
strings corresponds to different
vacua of the effective sigma model on the world sheet.
See also
the review paper \cite{SYrev} for details.

\section{World sheet theory}
\label{secWCP}
\setcounter{equation}{0}

In this section we   briefly describe the world-sheet low-energy sigma models on the non-Abelian strings
 at large and small $\xi$. Non-Abelian strings in \ntwo SQCD with $N_f=N$
where first found and studied in \cite{HT1,ABEKY,SYmon,HT2}. Then we discuss
how the bulk duality translates into the world-sheet duality~\cite{SYdual}.

\subsection {World sheet theory at large \boldmath{$\xi$}}
\label{strings}

To warm up, we start from $N_f=N$.
The Abelian $Z_N$-string solutions break the  SU$(N)_{C+F}$ global group. As a result,
the non-Abelian strings have orientational zero modes associated with rotations of their color
flux inside the non-Abelian SU($N$ group.  
The global group is broken on the $Z_N$ string solution down to
${\rm SU}(N-1)\times {\rm U}(1)$.
Hence,
the moduli space of the non-Abelian string is described by the coset space
\beq
\frac{{\rm SU}(N)}{{\rm SU}(N-1)\times {\rm U}(1)}\sim CP(N-1)\,,
\label{modulispace}
\eeq
in addition to $C$ spanned by the translational modes. The translational moduli totally decouple.
They are sterile free fields, and we can forget about them.
Therefore, the
low-energy effective theory on the   non-Abelian string
is the two-dimensional \ntwo   CP$(N-1)$ model \cite{HT1,ABEKY,SYmon,HT2}.

Now   we add ``extra" quark flavors with degenerate masses, increasing $N_f$ from $N$ up to a certain value
$N_f >N$.
The strings emerging in such theory are semilocal.
In particular, the string solutions on the Higgs branches (typical
for multiflavor theories) usually are not fixed-radius strings, but, rather,
possess radial moduli, aka size moduli, see    \cite{AchVas} for a comprehensive review of 
the Abelian semilocal strings. The transverse size of such a string is not fixed.

Non-Abelian semilocal strings in \ntwo SQCD with $N_f>N$ were studied in
\cite{HT1,HT2,SYsem,Jsem}. 
The orientational
zero modes of the semilocal non-Abelian string are parametrized by a complex vector $n^P$ ($P=1, ..., N$),
 while its $\tN=(N_f-N)$ size moduli are parametrized by a complex vector
$\rho^K$ ($K=N+1, ..., N_f$). The effective two-dimensional theory
which describes the internal dynamics of the non-Abelian semilocal string is
the \ntwot weighted  CP model on a ``toric" manifold, which includes both types of fields. The bosonic 
part of the action
in the gauged formulation (which assumes taking the limit $e^2\to\infty$)
has the form\,\footnote{Equation (\ref{wcp}) and similar expressions below are given in Euclidean notation.}
\beqn
&&S = \int d^2 x \left\{
 \left|\nabla_{\alpha} n^{P}\right|^2 
 +\left|\tilde{\nabla}_{\alpha} \rho^K\right|^2
 +\frac1{4e^2}F^2_{\alpha\beta} + \frac1{e^2}\,
\left|\pt_{\alpha}\sigma\right|^2
\right.
\nonumber\\[3mm]
&+&\left.
2\left|\sigma+\frac{m_P}{\sqrt{2}}\right|^2 \left|n^{P}\right|^2 
+ 2\left|\sigma+\frac{m_{K}}{\sqrt{2}}\right|^2\left|\rho^K\right|^2
+ \frac{e^2}{2} \left(|n^{P}|^2-|\rho^K|^2 -2\beta\right)^2
\right\},
\nonumber\\[4mm]
&& 
P=1,...,N\,,\qquad K=N+1,...,N_f\,.
\label{wcp}
\eeqn
The fields $n^{P}$ and $\rho^K$ have
charges  +1 and $-1$ with respect to the auxiliary U(1) gauge field;
hence, the corresponding  covariant derivatives in (\ref{wcp}) are defined as 
\beq
 \nabla_{\alpha}=\d_{\alpha}-iA_{\alpha}\,,\qquad 
\tilde{\nabla}_{\alpha}=\d_{\alpha}+iA_{\alpha}\,,
\label{covarder}
\eeq
  respectively.

If only charge $+1$ fields were present, in the limit
$e^2\to\infty$ we would get a conventional twisted-mass deformed
CP $(N-1)$ model.
The presence of  charge $-1$ fields $\rho^K$ converts the CP$(N-1)$
 target space   into that of the a weighted
CP$(N_f-1)$ model.
As in the CP$(N-1)$ model, small mass differences
$\left| m_A-m_B\right|$ lift orientational and size zero modes generating a shallow potential on the modular space.
The $D$-term condition
\beq
  |n^P|^2 - |\rho^K|^2=2\beta
\label{unitvec}
\eeq
is implemented in the limit $e^2\to\infty$. Moreover, in this limit
the gauge field $A_{\alpha}$  and its \ntwo bosonic superpartner $\sigma$ become
auxiliary and can be eliminated.

The  two-dimensional coupling constant $\beta$ is related to the four-dimensional
one as
\beq
\beta= \frac{2\pi}{g_2^2}\,.
\label{betag}
\eeq
This relation  is obtained  at the classical level 
\cite{ABEKY,SYmon}.
 In quantum theory
both couplings run. In particular, the  model (\ref{wcp}) is asymptotically free
\cite{W93} and develops its own scale $\Lambda_{\sigma}$. 
The ultraviolet cut-off in the sigma model on the string world sheet
is determined by  $g_2\sqrt{\xi}$. 
Equation~(\ref{betag}) relating the two- and four-dimensional couplings
is valid at this scale.
At this scale the four-dimensional coupling is given by (\ref{4coupling})
while the two-dimensional one
\beq
4\pi\beta(\xi) =
\left(N-\tN \right)\ln\,{\frac{g_2\sqrt{\xi}}{\Lambda_\sigma}}\gg 1 \,.
\label{2coupling}
\eeq
Then Eq.~(\ref{betag}) implies
\beq
\Lambda_{\sigma} =  \Lambda\, ,
\label{lambdasig}
\eeq
and from now on we will omit the subscript $\sigma$.
In the bulk, the running of the coupling constant is frozen at
$g_2\sqrt{\xi}$, because of the VEVs of
the squark fields.
The logarithmic evolution of the coupling constant in the string world-sheet theory 
continues uninterrupted below this point, with the same running law.
As a result, the dynamical scales of the bulk and world-sheet
theories turn out to be the same, much in the same way as in the $N_f=N$ theory \cite{SYmon}.

 We remind that the scale $g_2\sqrt{\xi}$ determines the ultraviolet cut-off 
in the sigma model on the
 string world sheet. Therefore we can consider the theory (\ref{wcp}) as an effective
theory on the string only at energies   below $g_2\sqrt{\xi}$. This is fulfilled if
\beq
\sqrt{\xi}\gg {\rm max}\left(|m_A-m_B|,\Lambda\right).
\label{cpcond}
\eeq

Summarizing,
if the quark mass differences are small, the
$(1 , ... ,\, N)$ vacuum of the original   U$(N)$ gauge theory  supports non-Abelian 
semilocal strings. Their internal dynamics is 
described by the effective two-dimensional low-energy \ntwot sigma model (\ref{wcp}).
The model has $N$ orientational moduli $n^P$ with the U(1) charge $+1$ and masses $m_P=\{m_1,...,m_N\}$,
plus
$\tN$ size moduli $\rho^K$, with the U(1) charge $-1$ and masses $(-m_K)=-\{m_{N+1},...,m_{N_f}\}$.

A final remark is in order here. The
strict semilocalality achieved at $\Delta m_{AB}=0$
destroys confinement of monopoles
\cite{EY,SYsem}. The reason is that the string  transverse size (determined by $\rho^K$'s )
can grow indefinitely.
When it becomes of the order of the distance $L$ between sources of the magnetic flux (the monopoles),
the linearly rising confining potential between these sources is replaced by a Coulomb-like
potential. To have confinement of monopoles  we should
lift the size zero modes keeping $\Delta m$ nonvanishing.
That's exactly what we will do, eventually sticking to (\ref{masssplit}),
preserving both confinement and the global symmetry.

\subsection{Dual world-sheet theory}
\label{dwsth}

The dual bulk U$(\tN)$ theory at small $\xi$ also supports
 non-Abelian semilocal strings. The $(1,...,N)$  vacuum  of the original theory 
(\ref{qvev}) transforms into the vacuum  (\ref{Dvev}) of the dual theory. Therefore, the internal 
string dynamics on the string world sheet
is described by a similar \ntwot sigma model. Now it has $\tN$ orientational moduli  with the 
U(1) charge $+1$ and masses $m_K=\{m_{N+1},...,m_{N_f}\}$. 
To make contact with (\ref{wcp}) we  call them $\trho^K$.
In addition,  it has $N$ size moduli  with the U(1) charge $-1$ and masses $(-m_P)=-\{m_1,...,m_N \}$. We 
refer to these  size moduli as 
$\tn^P$.

The bosonic part of the action of the world-sheet model 
in the gauged formulation (which assumes taking the 
limit $\tilde{e}^2\to\infty$) has the form
\beqn
S_{{\rm dual}} \! \! \!  &=& \! \! \!  \int d^2 x \left\{
 |\nabla_{\alpha} \trho^{K}|^2 +|\tilde{\nabla}_{\alpha} \tn^P|^2
 +\frac1{4e^2}F^2_{\alpha\beta} + \frac1{e^2}\,
|\pt_{\alpha}\sigma|^2
\right.
\nonumber\\[3mm]
&+&\! \! \! 
\left.
2\left|\sigma+\frac{m_P}{\sqrt{2}}\right|^2 \left|\tn^{P}\right|^2 
+ 2\left|\sigma+\frac{m_{K}}{\sqrt{2}}\right|^2\left|\trho^K\right|^2
+ \frac{e^2}{2} \left(|\trho^K|^2-|\tn^{P}|^2 -2\tilde{\beta}\right)^2
\right\},
\nonumber\\[3mm]
&& 
P=1,...,N\,,\qquad K=N+1,...,N_f\,,
\label{dcp}
\eeqn
where the covariant derivatives are defined in (\ref{covarder}).

We see that the roles of the orientational
and size moduli   interchange  in Eq.~(\ref{dcp})  compared with (\ref{wcp}).
As in the  model (\ref{wcp}), small mass differences
$\Delta m_{AB}$  lift orientational and size zero modes of the non-Abelian semilocal string generating a
 shallow potential on the moduli space.
Much in the same way as in the model (\ref{wcp}), the dual coupling constant $\tilde{\beta}$ is
determined by the bulk dual coupling $\tilde{g}_2^2$,
\beq
4\pi\tilde{\beta}(\xi) = \frac{8\pi^2}{\tilde{g}_2^2}(\xi)=
(N-\tN )\ln{\frac{\Lambda}{\tilde{g_2}\sqrt{\xi}}}\,,
\label{d2coupling}
\eeq
see Eqs.~(\ref{betag}) and  (\ref{2coupling}). 
The dual theory makes sense at $\tilde{g}_2 \sqrt\xi\ll \Lambda$ where $\tilde\beta$ is positive and
$$
4\pi\tilde{\beta}(\xi)\gg 1
$$
(weak coupling).

The bulk and world-sheet
dual theories  have identical $\beta$ 
functions, with the first coefficient $(\tN-N)<0$. They are both infrared (IR) free. As 
in the model (\ref{wcp}), the coincidence of the $\beta$ functions in the bulk
and world-sheet theories implies that the scale of the dual model (\ref{dcp}) is equal to that of the 
bulk theory,
$$\tilde{\Lambda}_{\sigma}=\Lambda\,,$$ 
cf. Eq.~(\ref{lambdasig}).
Comparing (\ref{d2coupling}) with (\ref{2coupling}) we see that 
\beq
\tilde{\beta}=-\beta\,.
\label{tbb}
\eeq
At $\xi\gg\Lambda^2$ 
the original theory is at weak coupling, and $\beta$ is positive. Analytically continuing to the domain
$\xi\ll\Lambda^2$, we formally make $\beta$ negative, which signals, of course, that the low-energy description
in terms of the original model is inappropriate. At the same time, $\tilde\beta$ satisfying 
(\ref{tbb}) becomes positive, and the dual model assumes the role of the legitimate low-energy description
(at weak coupling). A direct inspection of 
the dual theory action (\ref{dcp}) shows that 
the dual theory can be interpreted as a continuation of the sigma model (\ref{wcp})
to  negative values of the coupling constant $\beta$.

Both  world-sheet theories (\ref{wcp}) and (\ref{dcp}) give the {\em effective low-energy}
 descriptions of  string dynamics valid  at the energy scale below $g_2\sqrt{\xi}$.

Let us note that
  the world-sheet duality between
two-dimensional sigma models (\ref{wcp}) and (\ref{dcp}) was previously noted in Ref.~\cite{Jsem}. 
In this paper two bulk theories,  with the U$(N)$ and U$(\tN)$ gauge groups, were considered
(these theories were referred to as a dual pair in \cite{Jsem}). Two-dimensional 
sigma  models (\ref{wcp}) and (\ref{dcp})
were presented as  effective low-energy descriptions of the non-Abelian strings for 
these two bulk theories.

\section{Semiclassical description of the world-sheet theories}
\label{secclasslim}
\setcounter{equation}{0}

At $N<N_f<2N$ the original  model (\ref{wcp}) is asymptotically free, see (\ref{2coupling}). Its coupling $\beta$
continues running below $g_2\sqrt\xi$ until it stops at the scale of the mass differences
$\Delta m_{AB}$.
 If all mass differences are large, 
$|\Delta m_{AB}| \gg \Lambda$, the model is at weak
coupling. From (\ref{wcp}) we see that the model has $N$ vacua (i.e. $N$
strings from the standpoint of the bulk theory),
\beq
 \sqrt{2}\sigma=-m_{P_0},\qquad |n^{P_0}|^2=2\beta\,,\qquad n^{P\neq P_0}=\rho^K=0\,,
\label{classvac}
\eeq
where $P_0=1, ..., N$.

In each vacuum there are $2(N_f-1)$ elementary excitations, counting
real degrees of
freedom. The action (\ref{wcp}) contains $N$ complex fields
$n^P$ and $\tN$ complex fields $\rho^K$.
The  phase of $n^{P_0}$ is eaten by the Higgs mechanism.
The condition $|n^{P_0}|^2 = 2\beta$ eliminates one extra field.
The physical masses of the elementary excitations
\beq
M_A = |m_A-m_{P_0}|\,,\qquad A\neq P_0\,.
\label{elmass}
\eeq
In addition to the elementary excitations, there are kinks (domain ``walls" which are particles in two
dimensions) interpolating between different vacua. 
Their masses scale as
\beq
M^{\rm kink} \sim \beta\, M_A \,.
\label{kinkmasscl}
\eeq
The kinks  are much  heavier than elementary
excitations at weak coupling.\footnote{ Note that they have nothing to do
with Witten's $n$ solitons \cite{W79} identified as the $n^P$ fields at
strong coupling. In the next section we present a general formula for the kink spectrum
outside CMS (at weak coupling).}

Now we pass to the dual world-sheet
theory (\ref{dcp}). It is {\em not} asymptotically free (rather, IR-free) and, therefore,
is at weak coupling at small mass differences, $|m_{AB}|\ll \Lambda$.
From (\ref{dcp}) we see that this model has $\tN$ vacua 
\beq
 \sqrt{2}\sigma=-m_{K_0},\qquad |\rho^{K_0}|^2=2\tilde{\beta}\,,
\qquad n^{P}=\rho^{K\neq K_0}=0,
\label{classvacd}
\eeq
where $K_0=N+1,...,N_f$.
In each vacuum there are $2(N_f-1)$ elementary excitations
with the physical masses
\beq
M_A = |m_A-m_{K_0}|\,,\qquad A\neq K_0\,.
\label{elmassd}
\eeq
The dual model   has kinks too; their masses scale as (\ref{kinkmasscl}). 

It is important to understand 
that the dual theory (\ref{dcp}) is not asymptotically free
at energies much larger than the mass differences. At energies   smaller than some
mass differences certain fields decouple, and the theory may or may not become
asymptotically free. Keeping in mind the desired limit (\ref{masssplit})  we will consider 
the following mass choice  in the dual theory
\beq
m_P\sim m_{P'}, \qquad m_K\sim m_{K'}, \qquad m_P-m_K\sim\Delta m
\label{massdef}
\eeq
where $P,P'=1,...,N$ and $K,K'=N+1,...,N_f$. Moreover, we will often consider
the mass hierarchy
\beq
|\Delta m_{PP'}|\sim |\Delta m_{KK'}|\ll |\Delta m|\ll \Lambda, 
\label{masshier}
\eeq
where $\Delta m_{PP'}=m_P-m_{P'}$ and $\Delta m_{KK'}=m_K-m_{K'}$.

\vspace{1mm}

Clearly, the dual model is not asymptotically free only if 
the mass differences
$\Delta m_{KK'}$ are not too small.
If we take 
\beq
|\Delta m_{KK'}| \ll  |\Delta m|\ll\Lambda
\label{dop1}
\eeq
 the model becomes asymptotically free below $|\Delta m|$.
In fact, 
the model then  reduces to the CP$(\tN-1)$ model with an
effective scale 
\beq
 \tilde{\Lambda}_{\rm LE}^{\tilde N}\equiv \frac{(\Delta m)^N}{\Lambda^{N-\tN}}\,,\qquad
 \tilde{\Lambda}_{\rm LE}\ll|\Delta m|\,.
\label{duallambdale}
\eeq
 In particular, if $|\Delta m_{KK'}|\lsim \tilde{\Lambda}_{LE}$, descending down to $\tilde{\Lambda}_{LE}$ we
 enter (more exactly, the dual CP$(\tN-1)$ enters)  the strong coupling regime.

Thus, there are {\em two} strong coupling regimes in the dual model. One is at large mass differences
$|m_{AB}|\gg \Lambda$ where the original model (\ref{wcp})   is
at weak coupling and provides an adequate description, while the other is at very small mass differences $\Delta m_{KK'}\lsim \tilde{\Lambda}_{LE}$ where the dual model effectively reduces to the strongly coupled 
CP$(\tN-1)$ model.

\section{Exact superpotential}
\label{secsup}
\setcounter{equation}{0}

The \cpn models are known to be described by an exact superpotential \cite{AdDVecSal,ChVa,W93,Dorey} of
the  Veneziano-Yankielowicz  type \cite{VYan}.
This superpotential was generalized to the case of the weighted CP models in 
\cite{HaHo,DoHoTo}. In this section we will briefly outline  this method.
Integrating out the fields $n^P$ and $\rho^K$  we can describe
 the original model  (\ref{wcp}) by the following
exact twisted superpotential:
\beqn
 {\cal W}_{\rm eff}
 & =& 
\frac1{4\pi}\sum_{P=1}^N\,
\left(\sqrt{2}\,\Sigma+{m}_P\right)
\,\ln{\frac{\sqrt{2}\,\Sigma+{m}_P}{\Lambda}}
\nonumber\\[3mm]
&-& 
\frac1{4\pi}\sum_{K=N+1}^{N_F}\,
\left(\sqrt{2}\,\Sigma+{m}_K\right)
\,\ln{\frac{\sqrt{2}\,\Sigma+{m}_K}{\Lambda}} 
\nonumber\\[3mm]
&-& \frac{N-\tN}{4\pi} \,\sqrt{2}\,\Sigma\, ,
\label{2Dsup}
\eeqn
where $\Sigma$ is a twisted superfield \cite{W93} (with $\sigma$ being its lowest scalar
component). 
Minimizing this superpotential with 
respect to $\sigma$ we get the vacuum field formula,
\beq
\prod_{P=1}^N(\sqrt{2}\,\sigma+{m}_P)
=\Lambda^{(N-\tN)}\,\prod_{K=N+1}^{N_f}(\sqrt{2}\,\sigma+{m}_K)\,.
\label{sigmaeq}
\eeq
Note that the roots of this equation coincide with the double roots of the Seiberg--Witten curve  of 
the bulk theory \cite{Dorey,DoHoTo}.  This is, of course, a manifestation of coincidence
of the Seiberg--Witten solution of the bulk theory with the exact solution of (\ref{wcp})
given by the superpotential (\ref{2Dsup}). As was mentioned in Sec.~\ref{intro},
this coincidence was observed in \cite{Dorey,DoHoTo} and   explained later 
in \cite{SYmon,HT2}.

Now, let us consider the effective superpotential of the dual world-sheet theory (\ref{dcp}).
It has the form
\beqn
 \widetilde{\mathcal W} _{\rm eff} 
 &= &
\frac1{4\pi}\sum_{K=N+1}^{N_f}\,
\left(\sqrt{2}\,\Sigma+{m}_K\right)
\,\ln{\frac{\sqrt{2}\,\Sigma+{m}_K}{\Lambda}}
\nonumber\\[3mm]
& -& 
\frac1{4\pi}\sum_{P=1}^{N}\,
\left(\sqrt{2}\,\Sigma+{m}_P\right)
\,\ln{\frac{\sqrt{2}\,\Sigma+{m}_P}{\Lambda}}
\nonumber\\[3mm]
 &-& \frac{\tN-N}{4\pi} \,\sqrt{2}\,\Sigma \, .
\label{2Ddsup}
\eeqn
We see that it coincides with the superpotential (\ref{2Dsup})  up to a sign. Clearly, both, the
root equations  and the BPS spectra,  are the same for the two sigma
models, as was expected \cite{SYdual}. 

Although classically the dual pair of the weighted CP models at hand are given by different
actions (\ref{wcp}) and (\ref{dcp}), in the quantum regime they reduce to the one and the same theory.
This is, of course, expected. Classically the couplings of both theories are determined
by the ultraviolet (UV) cut-off scale $\sqrt{\xi}$, see (\ref{2coupling}) and (\ref{d2coupling}).
However, in quantum theory these couplings run and, in fact, are determined by the mass differences.
Therefore, if $|\Delta m_{AB}|\gsim \Lambda$, the coupling $\beta$ is positive and we use
the original theory (\ref{wcp}). 
On the other hand, If   $|\Delta m_{AB}|\lsim \Lambda$, the coupling
$\beta$ becomes negative,  we use the dual theory (\ref{dcp}) which has positive
$\tilde{\beta}$, see Sec.~\ref{secclasslim}.
The bulk FI parameter $\xi$ no longer plays a role. Only the values of the
mass differences matter. 

\vspace{1mm}

It is instructive to summarize the situation.
The theory has three distinct regimes, namely,
\vspace{1mm}

 (i) The weak coupling domain in the original description at large mass differences, 
\beq
|\Delta m_{AB}|\gg \Lambda;
\label{largemasses}
\eeq

(ii)  The mixed regime in the dual description at intermediate masses,
\beq
\tilde{\Lambda}_{\rm LE} \ll |\Delta m_{AB}|\ll \Lambda\,,
\label{intermasses}
\eeq
where all mass differences above  are assumed to be of the same order. 
 Certain vacua (namely, $\tilde N$ vacua)  are at  weak coupling
and can be seen classically, see (\ref{classvacd}), while $N-\tilde N$ other vacua are at strong coupling.
 If, instead, we 
assume the  mass hierarchy (\ref{masshier}) then in order to keep $\tN$ vacua at weak coupling
we have to impose the condition 
\beq
|\Delta m_{KK'}|\gg \tilde{\Lambda}_{LE}\,,
\label{dop2}
\eeq
 see (\ref{duallambdale}).
This is the reason why we call this region ``intermediate mass" domain.

\vspace{1mm}

(iii) The strong coupling regime in the dual description at hierarchically small masses, 
\beq
|\Delta m_{PP'}|\sim |\Delta m_{KK'}|\lsim \tilde{\Lambda}_{LE}|\ll |\Delta m|\ll \Lambda\,.
\label{smallmasses}
\eeq

\vspace{3mm}

The  masses of the BPS kinks interpolating between the
vacua $\sigma_{I}$ and $\sigma_{J}$ are given  by the appropriate 
differences of the superpotential (\ref{2Dsup}) calculated at distinct roots \cite{HaHo,Dorey,DoHoTo},
\beqn
M^{\rm BPS}_{I J}
&=&
2\left|{\cal W}_{\rm eff}(\sigma_{J})-{\cal W}_{\rm eff}(\sigma_{I})\right|
\nonumber\\
&=&
\left|\frac{N-\tN}{2\pi} \,\sqrt{2}(\sigma_{J}-\sigma_{I})
- \frac1{2\pi}\sum_{P=1}^{N}\,
{m}_P\,\ln{\frac{\sqrt{2}\,\sigma_{J}+{m}_P}{\sqrt{2}\,\sigma_{I}+{m}_P}}
\right.
\nonumber\\
&+&
\left.
\frac1{2\pi}\sum_{K=N+1}^{N_f}\,
{m}_K\,\ln{\frac{\sqrt{2}\,\sigma_{J}+{m}_K}{\sqrt{2}\,\sigma_{I}+{m}_K}}
\right|
\, .
\label{BPSmass}
\eeqn
The  masses obtained from (\ref{BPSmass}) were shown  \cite{SYmon,HT2}
to coincide with those of monopoles and dyons in the bulk theory. The latter are 
given by the period integrals of the Seiberg--Witten curve \cite{Dorey,DoHoTo}. 

Now we will consider the vacuum structure and the kink spectrum in more detail in 
two quasiclassical regions -- at large mass differences (the original theory) and 
at intermediate mass differences (the dual theory).

\subsection{Large \boldmath{$|\Delta m_{AB}|$}}
\label{seclargemasses}

Consider the vacuum structure of the theory (\ref{sigmaeq})
in the  weak coupling regime $|\Delta m_{AB}|\gg \Lambda$. In this domain the model has $N$ vacua
which in the leading 
(classical) approximation are determined  by Eq.~(\ref{classvac}). Equation (\ref{sigmaeq}) reproduces 
this vacuum structure.
Namely, VEVs of $\sigma$ in each of the $N$ vacua (say, at $P=P_0$) are given by the corresponding mass $m_{P_0}$,
plus a small correction,
\beq
\sqrt{2}\sigma_{P_0} \approx -m_{P_0} 
+\Lambda^{N-\tN}\,\frac{\prod_{K=N+1}^{N_f}({m}_K-m_{P_0})}{\prod_{P\neq P_0}({m}_P-m_{P_0})}
\,.
\label{sigmaclass}
\eeq
The spectrum of kinks is given by Eq.~(\ref{BPSmass}). To be more specific, let us 
consider the kinks interpolating between 
the neighboring vacua\,\footnote{If the mass parameters $m_P$ 
are randomly scattered in the complex plane, how one should define
the ``neighboring vacua"? In the regime under consideration, for all $P$ the vacuum values
$\sigma_P$ are close to $-m_P/\sqrt{2}$. Assume $\sigma_{P_0}$ is chosen.
Then the neighboring vacuum $\sigma_{P_0+1}$ is defined in such a way that the difference
$|m_{P_0} - m_{P_0+1}|$  is the smallest in the set $|\Delta m_{P_0P}|$.}  
$P_0$ and $\mbox{$(P_0+1)$}$. Then we have
\beq
m^{\rm kink}=|m_D^{P_0+1}-m_D^{P_0}|\approx
\left| \left(m_{P_0}-m_{P_0+1}\right)\,\frac{N-\tN}{2\pi}\,\ln{\frac{\bar\Delta m_{AB}}{\Lambda}}\right|,
\label{mD}
\eeq
where 
$\bar\Delta m_{AB}$
is a certain   average value of the mass differences (it depends holomorphically on all
mass differences in the problem).
 Here   we use (\ref{sigmaclass}).
If  all mass
differences are of the same order 
so is $\bar\Delta m_{AB}$, although $\bar\Delta m_{AB}$ does not coincide with any of
the individual mass differences. For a generic choice of the mass differences
$\bar\Delta m_{AB}$ has a nonvanishing phase.

 We see that in the logarithmic approximation the kink mass
is proportional to the mass difference $(m_{P_0}-m_{P_0+1})$ times the coupling constant
$\beta$, as one would expect at weak coupling, cf. Eq.~(\ref{kinkmasscl}).

This is not the end of the story, however. The   logarithmic functions
in (\ref{BPSmass}) are multivalued,  and we have to carefully choose their branches.
Each logarithm in (\ref{2Dsup}) can be written in the integral form as
\beq
\frac{m_{A}}{2\pi}\,\ln{\frac{\sqrt{2}\,\sigma_{P_0+1}+{m}_A}{\sqrt{2}\,\sigma_{P_0}+{m}_A}}
=\frac{m_{A}}{2\pi}\,\int_{\sigma_{P_0}}^{\sigma_{P_0+1}}\,
\frac{\sqrt{2}\,d\sigma}{\sqrt{2}\,\sigma +m_A}
\,,\label{integral}
\eeq
with the integration contour to be appropriately chosen. Distinct choices 
differ by pole contributions  
\beq
{\rm integer}\times im_A
\label{dop3}
\eeq
 for different $A$.
 These different mass predictions
for  the BPS states correspond to dyonic kinks. In addition to the topological charge,
the kinks can carry Noether charges with respect to
the the global group (\ref{c+f}) broken down to U(1)$^{N_f}$ by the mass differences.
This produces a whole family of dyonic kinks.\footnote{ They represent confined 
monopoles and dyons with the root-like electric charges in the bulk theory.}
We stress that all  these kinks with the imaginary part (\ref{dop3}) in the mass formula
interpolate between the same pair of vacua: $P_0$ and $(P_0+1)$. Our aim is
to count their number and calculate their masses.

Generically there are  way too many choices of the integration contours in (\ref{BPSmass}).
Not all of them are realized. Moreover, the kinks present in the quasiclassical domain
decay on CMS or form new bound states, cf. e.g. \cite{svz,OlmezS}. Therefore the quasiclassical
spectrum outside CMS and quantum spectrum inside CMS are different. 
We have to use an additional input on the structure of kink solutions to find out
the correct form of the BPS spectrum. In this section
we will summarize information on the classical spectrum while in the remainder of the
present  paper we will use
the mirror representation \cite{HoVa} of the model at hand to obtain the quantum spectrum.

The general formula for the BPS spectrum can be written as follows \cite{Dorey}:
\beq
M^{\rm BPS}=\left| \sum_I m_{D}^I T_I + i\sum_A m_A S_A\right|,
\label{TSmass}
\eeq
where the first term is a nonperturbative  contribution and $T_I$ is the topological charge 
$N$-vector, while the second term 
represents the dyonic (the Noether charge) ambiguity discussed above, with $S_A$ describing a global 
U(1) charge of the given BPS
state with respect to the U(1)$^{N_f}$ group.

The topological charge is given by
\beq
T_P=\delta_{PP_0+1}-\delta_{PP_0}
\label{topcharge}
\eeq
for kinks interpolating between the vacua $P_0$ and $(P_0+1)$, while $m_D$'s are
approximately given by the logarithmic terms in (\ref{mD}),
\beq
m_D^{P_0}\approx m_{P_0}\,\frac{N-\tN}{2\pi}\,\ln{\frac{\bar\Delta m_{AB}}{\Lambda}}.
\eeq
 Equation~(\ref{mD}) corresponds to
the kink with $S_A=0$.

At weak coupling the  BPS kinks can be studied using the classical solutions of the 
first-order equations. Each given kink solution breaks the global U(1)$^{N_f}$ group. Therefore, the 
kinks acquire zero modes associated with rotations in this internal group.
Quantization of the corresponding dynamics gives rise to dyonic kinks which carry global charges
$S_A$. This program was carried out for the \cpn model in \cite{Dorey} and 
 for the weighted CP model (\ref{wcp}) in \cite{DoHoTo}. The result is
\beq
S_P=sT_P\,,\qquad S_K=0
\label{classdyoncharge}
\eeq
for $P=1,...,N$ and $K=N+1,...N_f$, where $s$ is integer. Thus, at large $|\Delta m_{AB}|$ we have an 
infinite tower of the
dyonic kinks with masses
\beqn
M^{\rm kink}
&\approx&
 \left| \left(m_{P_0}-m_{P_0+1}\right)\right|
\nonumber\\[3mm]
&\times&
\left| \,\frac{N-\tN}{2\pi}\,\ln{\frac{\bar\Delta m_{AB}}{\Lambda}}
- is\right|.
\label{kinkspectrlm}
\eeqn
The expression in the second line under the sign of the absolute value
has both, real and imaginary parts. The real part is obtained from the logarithm by replacing
$\bar\Delta m_{AB}$ under the logarithm by $|\bar\Delta m_{AB}|$. The imaginary part
includes the phase of $\bar\Delta m_{AB}$, which, in principle, 
could be obtained for any given set of the mass differences,
but in practice this is hard to do for generic mass choices. In addition, the imaginary part
includes $is$, where $s$ is an integer (positive, negative or zero). When we change $s$,
we scan all possible values of the U(1) charge (i.e. go through the entire set of dyons). 

In addition to the above monopoles/dyons,
 in this domain of $\Delta m_{AB}$ there are   elementary excitations, see Eq. (\ref{elmass}).
These excitations are BPS-saturated too and can be described by the general formula (\ref{TSmass})
with $T=0$ and $S_A=\delta_{AB}-\delta_{A P_0}$ for any $B=1, ..., N_f$ in 
the  $P_0$-vacuum \footnote{ The actual kink spectrum at weak coupling is more complicated
than the one in (\ref{kinkspectrlm}). the kink states from the tower (\ref{kinkspectrlm}) can form
bound states with different elementary states in certain special domains of the mass
parameters \cite{DoHoTo,LeeYi}.}.

The above spectrum changes upon passing through CMS. In particular, we will see  
that elementary excitations do not exist inside CMS. All excitations that survive inside CMS are 
the kinks
with nonvanishing topological charges. This is a two-dimensional counterpart of the 
bulk picture: the quarks and gauge bosons decay inside the strong coupling domain giving rise to the
monopole-antimonopole pairs, see Sec.~\ref{secbulk}.

\subsection{Intermediate masses}
\label{secintmass}

Now  let is consider the domain of intermediate mass differences, see Eq.~(\ref{intermasses}). Then
 Eq.~(\ref{sigmaeq}) has $(N-\tN)$ solutions with
\beq
\sqrt{2}\sigma=\Lambda\,\exp\left({\frac{2\pi i}{N-\tN}\,l}\right),\qquad l=0,...,(N-\tN-1).
\label{Lvac}
\eeq
We will refer to these vacua as the $\Lambda$-vacua. They are at strong coupling, and will be studied later.
In this section we consider other $\tN$ vacua, which are 
at weak coupling and seen classically in the dual
theory, see Eq.~(\ref{classvacd}).
For these vacua Eq.~(\ref{sigmaeq}) gives
\beq
\sqrt{2}\sigma_{K_0} \approx -m_{K_0} 
+\frac{1}{\Lambda^{N-\tN}}\,\frac{\prod_{P=1}^{N}({m}_P-m_{K_0})}{\prod_{K\neq K_0}({m}_K-m_{K_0})}
\,,
\label{sigmaclassd}
\eeq
where $K_0=N+1,...,N_f$. These $\tN$ vacua will be referred to as the {\em zero-vacua} since 
in these vacua, with
small masses, the $\sigma$ vacuum expectation values  are much smaller than in the
$\Lambda$-vacua.

Substituting this in Eq.~(\ref{BPSmass}) we get  the  spectrum of the kinks interpolating
between the neighboring vacua
$K_0$ and $K_0+1$,
\beq
M^{\rm kink}\approx
\left| \left(m_{K_0}-m_{K_0+1}\right)\,\frac{N-\tN}{2\pi}\,\ln\frac{\Lambda}{\bar\Delta m_{AB}}
+ is\left(m_{K_0}-m_{K_0+1}\right)\right|,
\label{kinkspectrim}
\eeq
where we take into account the  U(1) charges parallelizing the derivation of
Eq.~(\ref{kinkspectrlm}) and  applying the quantization
procedure of \cite{Dorey,DoHoTo} to the dual theory (\ref{dcp}). 
As previously,
all mass differences are assumed to be of the same order.

If, instead, we consider a stricter  mass hierarchy (\ref{masshier}) (still requiring that
we are at   weak coupling $|\Delta m_{KK'}|\gg \tilde{\Lambda}_{LE}$) then Eq.~(\ref{kinkspectrim})
must be modified. With this stricter hierarchy the product in the numerator of the
second term in (\ref{sigmaclassd}) reduces to $(\Delta m)^N$ to form $\tilde{\Lambda}_{LE}$,
and the kink spectrum takes the form
\beq
M^{\rm kink}\approx
\left| \left(m_{K_0}-m_{K_0+1}\right)\,\frac{\tN}{2\pi}\,
\ln\frac{ \bar\Delta m_{KK'}}{\tilde{\Lambda}_{LE}}
+ is\left(m_{K_0}-m_{K_0+1}\right)\right|,
\label{kinkspectrim2}
\eeq
where $\tilde{\Lambda}_{LE}$ is given by (\ref{duallambdale}). This is just the kink spectrum
in the CP$(\tN-1)$ model at weak coupling.

In addition to the $T\neq 0$ kinks,
there are elementary excitations with masses given in Eq.~(\ref{elmassd}).
They correspond to $T=0$ and $S_A=\delta_{AB}-\delta_{A K_0}$ for any $B=1,...,N_f$ in the $K_0$-vacuum
in (\ref{TSmass}).

Confronting Eqs.~(\ref{kinkspectrlm}) and (\ref{kinkspectrim}) 
we see that the kinks 
have different Noether charges in the domains of large and intermediate mass differences.
At large masses they have charges with respect to the first $N$ factors of the 
global U(1)$^{N_f}$ group, while the kinks at the intermediate masses are charged with respect to
the last $\tN$ factors (this would correspond to SU$(N)$ and SU$(\tN)$ factors
of the global group (\ref{c+f}) in the limit (\ref{masssplit})). Therefore, they are,
in fact, absolutely  distinct states. 
The BPS states decay/form new bound states upon passing from one domain to
another. The restructuring happens on CMS which are surfaces located at $|\Delta m_{AB}|\sim \Lambda$
in the mass parameter space.\footnote{Of course, this restructuring
is a reflection of the same phenomenon in 
the bulk
theory, see Sec.~\ref{secbulk}.} As we will see shortly,
 in the weighted CP model at hand we 
have  another set of CMS at much smaller mass differences $|\Delta m_{KK'}|\sim \tilde{\Lambda}_{LE}$.
This additional CMS separates the domain of intermediate masses from 
that at strong coupling, see (\ref{smallmasses}).

\section{Mirror description}
\label{secmirror}
\setcounter{equation}{0}

Now we turn to the study of the quantum BPS spectrum inside CMS. We will determine 
the BPS spectrum in the $\Lambda$-vacua (\ref{Lvac}) at intermediate  and small masses,
 as well as the spectrum in the zero-vacua in the strong coupling domain at
hierarchically small masses (\ref{smallmasses}). 

\subsection{Mirror superpotential}

To this end we will rely on the mirror
formulation \cite{HoVa} of the  weighted CP model (\ref{wcp}). In this formulation one
describes the CP model as a Coulomb gas of instantons (see \cite{FFS} where it was first 
done in the nonsupersymmetric CP(1) model). In supersymmetric setting this description leads
to an affine Toda theory with an exact superpotential. The exact mirror superpotentials
were found for the \ntwot \cpn model and its various generalizations with toric
target spaces in \cite{HoVa}. For the model (\ref{wcp}) the mirror superpotential has the form
\beq
W_{\rm mirror}=-\frac{\Lambda}{4\pi}\left\{\sum_{P=1}^{N} X_P -\sum_{K=N+1}^{N_f} Y_K
-\sum_{P=1}^{N} \frac{m_P}{\Lambda}\,\ln{X_P}+\sum_{K=N+1}^{N_f}\frac{m_K}{\Lambda}\,\ln{Y_K}
\right\}
\label{mirror}
\eeq
supplemented by the constraint
\beq
\prod_{P=1}^{N} X_P=\prod_{K=N+1}^{N_f} Y_K.
\label{constraint}
\eeq
This representation can be checked by a straightforward calculation. Indeed,  add the term 
\beq
\frac{\Lambda}{4\pi} \,\sqrt{2}\Sigma\left(\sum_{P=1}^{N} \ln{X_P}-\sum_{K=N+1}^{N_f} \ln{Y_K}
\right)
\eeq
to the superpotential (\ref{mirror}), which takes into account the constraint
(\ref{constraint}). Here $\Sigma$ plays a role of the Lagrange multiplier.
Then integrate over $X_P$ and $Y_K$ ignoring their kinetic terms. In this way one arrives at
\beq
X_P=\frac1{\Lambda}\left(\sqrt{2}\,\sigma+{m}_P\right),
\qquad Y_K=\frac1{\Lambda}\left(\sqrt{2}\,\sigma+{m}_K\right).
\label{XandY}
\eeq
Substituting (\ref{XandY}) back into (\ref{mirror}) one gets the superpotential 
(\ref{2Dsup}).

Clearly for the dual model (\ref{dcp}) the mirror superpotential coincides
with that in (\ref{mirror}) up to an (irrelevant) sign.

Below we will
find the critical points of  the superpotential (\ref{mirror}) and discuss 
the vacuum structure of the model in the mirror representation. Since our goal is
to study the domain of intermediate or hierarchically small masses, see (\ref{intermasses})
and (\ref{smallmasses}), respectively, we will assume that 
\beq
|\Delta m_{AB}|\ll \Lambda
\eeq 
 and keep only terms linear  in $|\Delta m_{AB}|/ \Lambda$. As a warm-up exercise 
 we will start with the 
\cpn model.

\subsection{\boldmath{\cpn} model}

For \cpn model $\tN=0$ and the superpotential  (\ref{mirror}) reduces to
\beq
W_{\rm mirror}^{{\rm CP}(N-1)}=-\frac{\Lambda}{4\pi}\left\{\sum_{P=1}^{N} X_P 
-\sum_{P=1}^{N} \frac{m_P}{\Lambda}\,\ln{X_P}\,,
\right\}
\label{mirrorcp}
\eeq
while the  constraint (\ref{constraint}) reads
\beq
\prod_{P=1}^{N} X_P=1\,.
\label{constraintcp}
\eeq
Expressing, say $X_1$ in terms of   $X_P$ with $P=2,3, ..., N$ by virtue of this constraint
and substituting the result in (\ref{mirrorcp}),  we get the 
 vacuum equations,
\beq
X_P=X_1+\frac{  m_{P}-m_1}{\Lambda}  = X_1+\frac{\Delta m_{P1}}{\Lambda}\,.\qquad P=2, ..., N\,.
\label{vaceqcp}
\eeq
Substituting this in (\ref{constraintcp}) we obtain $X_1$. This procedure
leads us to the following VEV's of the $X_P$  fields: 
\beq
X_P\approx \exp{\left(\frac{2\pi i}{N}\,l\right)} + \frac{1}{\Lambda}\left(
m_P-m\right)\,,\qquad \forall P\,,
\label{xvevcp}
\eeq
where we ignore quadratic in mass
differences terms, 
\beq
m\equiv\frac1N\sum_{P=1}^{N} m_{P}\,,
\label{m}
\eeq
and 
each of the $N$ vacua of the model is labeled by a value of $l$, namely
 $l=0$ in the first vacuum, $l=1$ in the second, and so on till we arrive at $l=(N-1)$.

\subsection{\boldmath{$\Lambda$}-vacua}
\label{lambdava}

Now we turn to the strong coupling vacua in  the weighted CP model at intermediate 
or small masses
(see Eq.~(\ref{smallmasses})), using 
the mirror representation (\ref{mirror}). 
These vacua were defined as $\Lambda$-vacua in Sec.~\ref{secintmass}, Eq.~(\ref{Lvac}).

Again expressing, say, $X_1$
 in terms of other fields by virtue of the constraint (\ref{constraint}) we get the vacuum equations
\beq
X_P=X_1+\frac{\Delta m_{P1}}{\Lambda},
\qquad Y_K=X_1+\frac{\Delta m_{K1}}{\Lambda}\, .
\label{vaceqL}
\eeq
Substituting this in the constraint (\ref{constraint}) and resolving for $X_1$ we get
the following VEVs:
\beqn
X_P
&\approx &
\exp{\left(\frac{2\pi i}{N-\tN}\,l\right)} + \frac{1}{\Lambda}\left(
m_P-\hat{m}\right),\quad P=1,...,N,
\nonumber\\[3mm]
Y_K
&\approx &
\exp{\left(\frac{2\pi i}{N-\tN}\,l\right)} + \frac{1}{\Lambda}\left(
m_K-\hat{m}\right),\quad K=N+1,...,N_f\,,
\nonumber\\[3mm]
l&=&0, ..., (N-\tN-1)
\label{vevL}
\eeqn
for $(N-\tN)$ vacua  of the theory. Here 
\beq
\hat{m}\equiv\frac1{N-\tN}\left(\sum_{P=1}^{N} m_{P}-\sum_{K=N+1}^{N_f}m_{K}\right)\,.
\label{hatm}
\eeq
As in (\ref{xvevcp}), in Eq.~(\ref{vevL}) we neglect
the  quadratic in mass difference terms, cubic, and so on.

\subsection{Zero-vacua}
\label{zerovacu}

Now consider other $\tN$ vacua of the model (\ref{dcp}) (which were termed zero-vacua
in Sec.~\ref{secintmass}) using the mirror description (\ref{mirror}). At intermediate masses,
VEVs of $\sigma$ are given, in the classical approximation,  by the mass parameters $m_K$  in the dual 
theory, see (\ref{classvacd}). The corrections are given by (\ref{sigmaclassd}).
Since the mirror representation is particularly useful at strong coupling, in this section we will
focus on a very small hierarchical region of the mass parameters (\ref{smallmasses}).

It is convenient to express one of $Y_K$'s, say, $Y_{N_f}$ in terms of other fields
using (\ref{constraint}). Then  vacuum equations the take the form
\beq
X_P=Y_{N_f}+\frac{\Delta m_{PN_f}}{\Lambda}\,,
\qquad Y_K=Y_{N_f}+\frac{\Delta m_{KN_f}}{\Lambda}\,.
\label{vaceqz}
\eeq
From the first equation   we see that, given the  hierarchical masses (\ref{smallmasses}),
all $X_P$ fields are equal to each other to the leading order,
\beq
X_P^{(0)}\approx \frac{\Delta m}{\Lambda}\,, \qquad P=1, ..., N\,.
\label{Xvevzero}
\eeq
With these $X_P$'s the constraint (\ref{constraint}) takes the form
\beq
\prod_{K=N+1}^{N_f} Y_K=\left(\frac{\tilde{\Lambda}_{LE}}{\Lambda}\right)^{\tN},
\label{constraintY}
\eeq
where $\tilde{\Lambda}_{LE}$ is the  scale of the effective low energy CP$(\tN-1)$ model  
 (\ref{duallambdale}). 
Substituting the fields $Y_K$ from
the second equation in  (\ref{vaceqz}) we get
\beq
Y_K\approx \frac{\tilde{\Lambda}_{LE}}{\Lambda}\,\left\{
\exp{\left(\frac{2\pi i}{\tN}\,l\right)} + \frac{1}{\tilde{\Lambda}_{LE}}\left(
m_K-\tilde{m}\right)\right\},
 \label{Yvev}
\eeq
where $l=0,...,\tN-1$,
\beq
\tilde{m}\equiv\frac1{\tN}\sum_{K=N+1}^{N_f}m_{K}\,.
\label{tildem}
\eeq
As usual,  we neglect quadratic, cubic, etc.  mass-difference  terms.

Finally, we are ready to solve the vacuum equations.
Substituting (\ref{Yvev}) in the first expression in (\ref{vaceqz}) 
we get  $O(\tilde{\Lambda}_{LE}/\Lambda )$ corrections to (\ref{Xvevzero}),
\beq
X_P\approx \frac{1}{\Lambda}(m_P-\tilde m)+\frac{\tilde{\Lambda}_{LE}}{\Lambda}\,
\exp{\left(\frac{2\pi i}{\tN}\,l\right)}, \qquad l=0,...,\tN-1\,.
\label{Xvev}
\eeq
We see that there are exactly $\tN$ vacua with very small values of $Y_K$'s. The VEV structure 
 of $Y_K$'s reduces to that of the CP$(\tN-1)$ model with the scale 
 parameter $\tilde{\Lambda}_{LE}$,
see (\ref{xvevcp}). 

\section{Kinks inside CMS}
\label{seckink}
\setcounter{equation}{0}

In this section we use the mirror representation to find 
the kink spectrum inside CMS in the
weighted CP model on the string world sheet. First, we briefly
review the kink solutions and their spectrum \cite{HoVa}
in the $\mbox{\cpn}$ model   and only then turn to the weighted CP model
(\ref{wcp}).

\subsection{Kinks in the \boldmath{\cpn} model}
\label{seccpnkinks}

As was shown in \cite{HoVa},  in the strong coupling regime (inside CMS) the number
of kinks interpolating between the vacua $P$ and $P+k$ of \ntwot  supersym\-metric \cpn model is
\beq
\nu(N,k)=C_N^k\equiv\frac{N!}{k!(N-k)!}\, .
\eeq
In particular, the number of kinks interpolating between the neighboring vacua ($k=1$) 
is $N$, and they
form a fundamental representation of the SU$(N)$ group. They carry the minimal charge with respect to
the gauge U(1) and, therefore, were identified \cite{W79} 
as $n^P$ fields in terms of the original description, see (\ref{wcp}) for $\tN=0$.

\begin{figure}
\epsfxsize=8cm
\centerline{\epsfbox{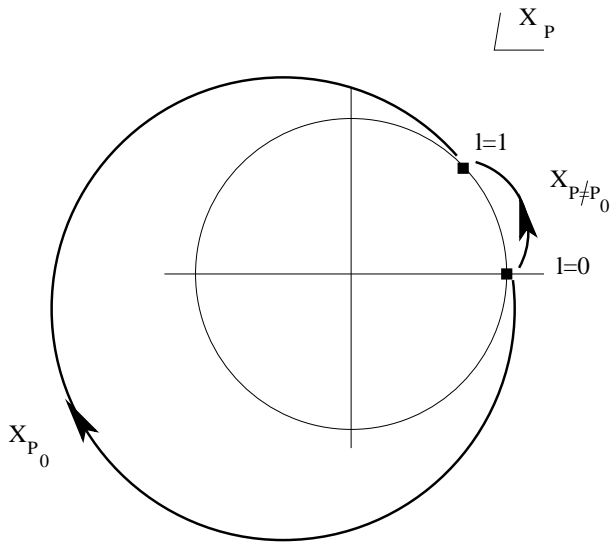}}
\caption{\small Windings of the fields $X_P$ for the kink interpolating between 
the $l=0$ and $l=1$ vacua.}
\label{figkinkcp}
\end{figure}

Consider a kink interpolating between the neighboring $l=0$ and $l=1$ vacua, see (\ref{xvevcp}).
The kink solution has the following structure \cite{HoVa}. All $X_P$'s start in the vacuum
 with $l=0$ and end in the vacuum with $l=1$. 
Moreover, all $X_P$ with $P\neq P_0$
(were $P_0$ is fixed for a given kink solution) are equal to each other and have
phases which wind by the angle $2\pi/N$ in the anti-clockwise direction. Then the constraint
(\ref{constraintcp}) ensures that  $X_{P_0}$ winds in the opposite (clockwise) direction 
by the angle $[-2\pi (N-1)/N]$, see Fig.~\ref{figkinkcp}. (If one considers
nonelementary kinks
interpolating between non-neighboring vacua, say $l=0$ and $l=2$, then two fields $X_{P_0}$ and  
$X_{P_0^\prime}$ would have 
opposite windings with respect to all others, and so on \cite{HoVa}.)

The kink mass is given by
\beqn
M^{\rm BPS}
&=&
2\left|{\cal W}_{\rm mirror}^{{\rm CP}(N-1)}(l=1)-{\cal W}_{\rm mirror}^{{\rm CP}(N-1)}(l=0)\right|
\nonumber\\[3mm]
&\approx&
\left|\frac{N}{2\pi}\,\Lambda\left(e^{\frac{2\pi}{N}\,i}-1\right) +i\left(m_{P_0}-
m\right)\right|,
\label{cpspectr}
\eeqn
where we use (\ref{mirrorcp}) and neglect quadratic in mass differences terms.
The parameter $m$ is defined in Eq.~(\ref{m}).

If we look at the absolute values of the fields $X_P$ rather than at their phases,
we will see that, generically, their absolute values differ from unity. This is discussed
in Appendix A, cf. also \cite{RitzSV}.
The explicit profile functions of the kink solutions are irrelevant 
for determination of the kink spectrum,
since the latter is given by 
central charges -- 
the differences of the superpotential on the initial and
final vacua. The phases of $X_P$ are important, however, because logarithms in (\ref{mirrorcp})
are multivalued functions. In particular, the term $im_{P_0}$ appeared in (\ref{cpspectr})
because $X_{P_0}$ has the relative winding angle $(-2\pi)$ with respect to other $X_P$'s.

We see that we have exactly $N$ dyonic kinks associated for the given choice of $P_0$
and its neighbor.
(In addition, $P_0$ can be chosen arbitrarily from the set $P_0=1,...,N$).
The above dyonic kinks
have different charges with respect to 
the global U(1)$^{N}$ and are split at generic masses, but become degenerate in the equal mass limit.
Clearly, they form a fundamental representation of the global SU$(N)$ in this limit.
We stress again that all $N$ kinks here interpolate between two fixed vacua, $l=0$ and $l=1$.

The BPS spectrum inside CMS, see (\ref{cpspectr}), is very different from that
outside CMS, see (\ref{kinkspectrlm}) with $\tN=0$. The weak coupling spectrum has an 
infinite tower of dyonic kinks, all associated with the same mass difference
$(m_{P_0}-m_{P_0+1})$. Also, the weak coupling spectrum has elementary states
with $T_P=0$. The quantum spectrum has only a finite number of states ($N$), with masses
which depend on all mass differences present in the theory. Moreover, all these states
are topological (they are kinks), no elementary states are present. The majority of the BPS states
present at weak coupling (in particular, all elementary excitations) decay on CMS and are  
absent at strong coupling.

In conclusion we note that Eq.~(\ref{BPSmass}) is exact and,
in principle, can be used to calculate the BPS spectrum in any domain of the parameter space
of the theory. We can put $\tN=0$ in this formula (descending down to the \cpn model) and 
apply (\ref{BPSmass}) to the kinks which interpolate between two  vacua $l=0$ and $l=1$ with 
VEVs of $\sigma$ given by 
\beq
\sqrt{2}\sigma\approx \Lambda\,\exp\left({\frac{2\pi i}{N}\,l}\right),\qquad l=0,...,N
\label{cpvac}
\eeq
at small masses, cf. (\ref{Lvac}) with $\tN=0$. 

\begin{figure}
\epsfxsize=10cm
\centerline{\epsfbox{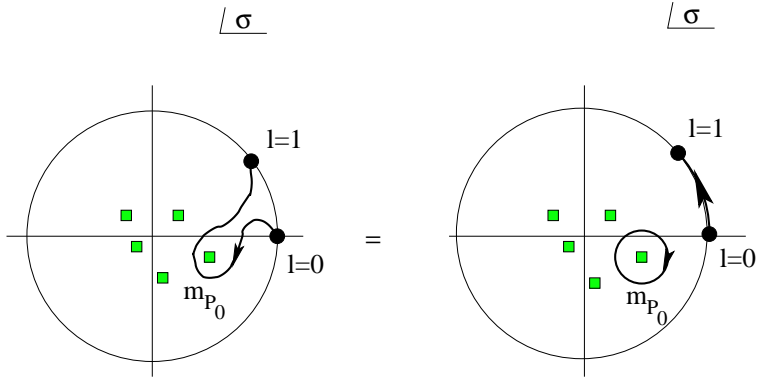}}
\caption{\small Integration contour in the $\sigma$ plane. Dots denote two vacua
$l=0$ and $l=1$, while filled squires denote poles located at $\sqrt{2}\sigma=-m_{P}$
for $P=1, ..., N$.}
\label{figcontourcp}
\end{figure}

Say, the main contribution in (\ref{cpspectr}) proportional to $\Lambda$
 comes from the first nonlogarithmic term in the second line in
(\ref{BPSmass}). Moreover, now the result in (\ref{cpspectr}) shows how  ambiguities
related to the choice of the logarithm branches  in (\ref{BPSmass}) should be
resolved at strong coupling. Namely, we get exactly the same BPS spectrum 
 as in (\ref{cpspectr}) from (\ref{BPSmass}) (with $\tN=0$) if we choose the 
integration contours in (\ref{integral}) as shown in Fig.~\ref{figcontourcp}.
For the $P_0$-th dyonic kink , the integration contour should pick up exactly one  pole contribution
located at $\sqrt{2}\sigma=-m_{P_0}$ in the clockwise direction. 
This  shows, in fact, how the kink solutions look  in terms of the field $\sigma$.

The above recipe was obtained in the  \cpn model in \cite{HaHo} using a
brane construction,
see also \cite{DoHoTo}. Here we confirm it directly in field theory 
using the mirror description of the model. See also \cite{RitzSV}.

\subsection{Kinks in the \boldmath{$\Lambda$}-vacua}
\label{kilva}

In this section  we work out a similar procedure to obtain the  BPS spectrum in 
the weighted CP model
(\ref{dcp}). We will focus on the   $\Lambda$-vacua 
 at intermediate or small masses, see (\ref{intermasses}) and  (\ref{smallmasses}).

Consider  kinks interpolating between the
neighboring $l=0$ and $l=1$ vacua (\ref{vevL}). 
Much in the same way as in the \cpn model (Sec.~\ref{seccpnkinks}), all $X_P$'s and $Y_K$'s are equal 
to each other and 
wind by the angle $2\pi/(N-\tN)$, except one field whose winding angle is determined by the 
constraint (\ref{constraint}). There are two different types of kinks depending on
the choice of the latter variable: $X_{P_0}$ or $Y_{K_0}$, ($P_0=1, ..., N$ and
$K_0=(N+1), ..., N_f$). We refer to these two types 
of solutions as to the $P$-  and $K$-kinks, respectively.

\subsubsection{\boldmath{$P$}-kinks}
\label{subsub1}

Solutions for $P$-kinks are very similar to those for kinks in the $\mbox{\cpn}$ model.
In a given $P$-kink the variable $X_{P_0}$ winds in the clockwise direction by
the angle $[-2\pi (N-\tN-1)/(N-\tN)]$, see Fig.~\ref{figkinkcp}.
All other fields, i.e.  $X_{P\neq P_0}$ and  $Y_K$, are equal to each other and   wind
counterclockwise by 
the angle $2\pi/(N-\tN)$.

The superpotential (\ref{mirror}) implies that the mass
of this kink is
\beq
M^{\rm BPS}_{P_{0}} \approx
\left|\frac{N-\tN}{2\pi}\,\Lambda\left(e^{\frac{2\pi}{N-\tN}\,i}-1\right) +i\left(m_{P_0}-
\hat{m}\right)\right|,
\label{LspectrP}
\eeq
where $\hat{m}$ is given in Eq.~(\ref{hatm}).

 We see that in the transition $l=0 \longrightarrow l=1$ 
 we have exactly $N$  kinks associated with the arbitrary choice of of the contour (with the loop around the pole 
 $P_0=1,...,N$).
They are split for generic masses, but become degenerate in the limit (\ref{masssplit})
we are interested in.
They form the  fundamental representation of the global SU$(N)$ in this limit.

\subsubsection{ \boldmath{$K$}-kinks}
\label{subsub2}

\begin{figure}
\epsfxsize=8cm
\centerline{\epsfbox{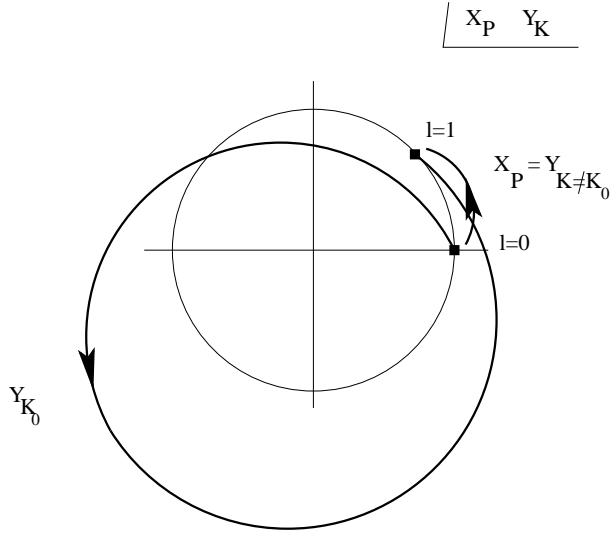}}
\caption{\small Windings of f
the fields $X_P$ and $Y_K$ for the $K$-kink interpolating between the
$l=0$ and $l=1$ $\Lambda$-vacua.}
\label{figkinkL}
\end{figure}

In the  $K$-kink solutions all fields $X_{P}$ and  $Y_{K\neq K_0}$ are equal to each other and have 
the winding  
 angle $2\pi/(N-\tN)$, while  the field $Y_{K_0}$ winds   in  the anticlockwise direction by
the angle $[2\pi (N-\tN+1)/(N-\tN)]$, see Fig.~\ref{figkinkL}, as dictated\,\footnote{ In fact,
for  these 
solutions to exist   $(N-\tN)$ is required to be large enough.
If $(N-\tN)$ is not large enough, the functions of $|X_P|$ and $|Y_K|$
develop singularities;  the singular solutions must be discarded, cf. Appendix A. The reason
behind this phenomenon  is that for 
$(N-\tN)$  not large enough the vacuum which is the closest neighbor to $l=0$   is, in fact,  
one of the zero-vacua (Sec.~\ref{zerovacu}) rather than  the $l=1$ $\Lambda$-vacuum.}
by the constraint
(\ref{constraint}). Thus, the $Y_{K_0}$ field has the relative winding   $+2\pi$ with respect
to all other fields.

\noindent
Substituting this 
information in the superpotential (\ref{mirror}) we obtain the $K$-kink mass,
\beq
M^{\rm BPS}_{K_0} \approx
\left|\frac{N-\tN}{2\pi}\,\Lambda\left(e^{\frac{2\pi}{N-\tN}\,i}-1\right) +i\left(m_{K_0}-
\hat{m}\right)\right|.
\label{LspectrK}
\eeq
Clearly, we have  $\tN$  kinks of this type due to the possibility of choosing the contour
encompassing any of the $\tN$ poles, 
 $K_0=(N+1), ..., N_f$.
They are split at generic masses, but become degenerate in the limit (\ref{masssplit}).
They form the  fundamental representation of the global SU$(\tN)$ group in this limit.

Altogether we have $N_f$ kinks interpolating between each pair of neighboring $\Lambda$-vacua.
They form the fundamental representation of the global group (\ref{c+fd}) in the 
limit (\ref{masssplit}). More exactly, they form the $(N,1)+(1,\tN)$ representation of this
group.

Much in the same way as in the \cpn model,  we can verify that Eq.~({\ref{BPSmass}) reproduces
this spectrum with the appropriate choice of the integration contours. Namely,
for the $P_0$-kink the contour in the $\sigma$ plane encircles the pole at 
$\sqrt{2}\sigma=-m_{P_0}$ in the clockwise direction, see Fig.~\ref{figcontourcp}.
 For the $K_0$-kink
the contour encircles the pole at 
$\sqrt{2}\sigma=-m_{K_0}$ in the anticlockwise direction.

\subsection{Kinks in the zero-vacua}
\label{seckinkZ}

Now we consider kinks in the
zero-vacua in the domain of 
small hierarchical masses (\ref{smallmasses}).
These vacua of the weighted CP model are most interesting since they correspond
to $\tN$ non-Abelian strings of the dual bulk theory. Clearly, the number of kinks
does not depend on which pair of neighboring vacua we pick up. Thus, we expect to have
altogether $N_f$ kinks, as was the case in the $\Lambda$-vacua. We check this explicitly below.

Much in the same way as in the $\Lambda$-vacua, the  kinks interpolating between 
the neighboring
$l=0$ and $l=1$ zero-vacua (see (\ref{Yvev})) fall into two categories: the $P$-
and $K$-kinks, respectively,  depending on the choice of the particular $X_{P_0}$ or $Y_{K_0}$ field
with an opposite winding with respect to other fields.

\subsubsection{\boldmath{$K$}-kinks}
\label{subsub3}

Let us start from the $K$-kinks. The kink solution looks very similar to that in the
CP$(\tN-1)$ model. All $Y_{K\neq K_0}$ fields are approximately equal to each other and
have the winding angles $2\pi/\tN$, while the $Y_{K_0}$ field has the winding angle $-2\pi(\tN-1)/\tN$,
see (\ref{Yvev}). Correction terms in $X_P$, proportional to $\tilde{\Lambda}_{LE}/\Lambda$
also all have the same windings by the angle $2\pi/\tN$, see (\ref{Xvev}).
This gives the following expression for the  mass of the $K_0$-kink:
\beq
M^{\rm BPS}_{K_0} \approx
\left|\frac{N-\tN}{2\pi}\,\tilde{\Lambda}_{LE}\left(e^{\frac{2\pi}{\tN}\,i}-1\right) -i\left(m_{K_0}-
\tilde{m}\right)\right|.
\label{ZspectrK}
\eeq
The factor $(N-\tN)$ in the first term appears from two first terms
in (\ref{mirror}) due to winding of both $X_P$ and 
$Y_K$ fields. The $im_{K_0}$ term is due to the relative winding $-2\pi$ of
the  $Y_{K_0}$ field.

We have  $\tN$  kinks of this type associated with the arbitrary choice of
 $K_0=(N+1), ..., N_f$.
They are split with generic masses, but become degenerate in the limit (\ref{masssplit}). In this limit they
form the  fundamental representation of the global SU$(\tN)$ group.

\subsubsection{\boldmath{$P$}-kinks}
\label{subsub4}

In this case  all $Y_K$ fields are approximately equal to each other and have 
the winding angles $2\pi/\tN$. The fields $X_{P\neq P_0}$ are all equal to $\Delta m/\Lambda$,
 to the leading order, and 
do {\em not} wind; however, they have windings $2\pi/\tN$ in the correction terms, see (\ref{Xvev}).
The field $X_{P_0}$ does wind. Its absolute value is $\Delta m/\Lambda$ . The winding angle is
$2\pi$, as enforced by the constraint (\ref{constraint}), see Fig.~\ref{figkinkz}. A more detailed
description of the kink solutions is presented in Appendix B. The windings above imply 
the following   mass of the $P_0$-kink:
\beqn
M^{\rm BPS}_{P_0} 
&\approx&
\left|\frac{N-\tN}{2\pi}\,\tilde{\Lambda}_{LE}\left(e^{\frac{2\pi}{\tN}\,i}-1\right) -i\left(m_{P_0}-\tilde{m}\right)\right|
\nonumber\\[4mm]
&=&
\left|-i\Delta m+\frac{N-\tN}{2\pi}\,\tilde{\Lambda}_{LE}\left(e^{\frac{2\pi}{\tN}\,i}-1\right) -i\left(m_{P_0}-m\right)\right|,
\label{ZspectrP}
\eeqn
where $m$ and $\tilde{m}$ are given by (\ref{m}) and (\ref{tildem}), while
$\Delta m=m-\tilde m$, see (\ref{massdef}). 
In the last expression the first term is the leading contribution, the second one is
a correction, while the third term accounts for still smaller splittings.

\begin{figure}
\epsfxsize=8cm
\centerline{\epsfbox{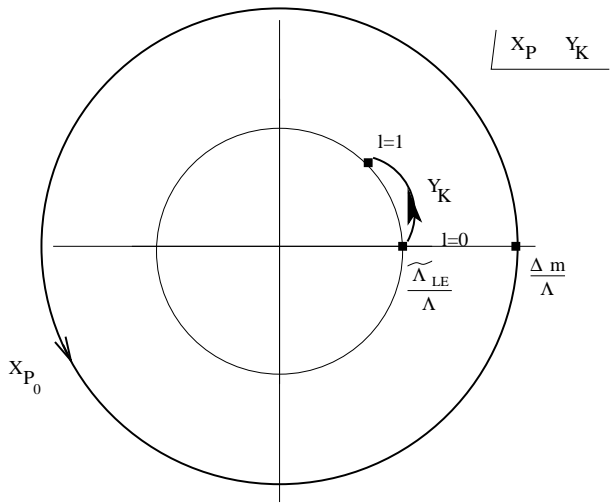}}
\caption{\small Windings of fields $X_{P_0}$ and $Y_K$ for the $P_0$-kink interpolating between $l=0$ and $l=1$ zero-vacua.}
\label{figkinkz}
\end{figure}

There are  $N$  kinks of this type associated with the arbitrary choice of
 $P_0=1,...,N$.
They are split at generic masses, but become degenerate in the limit (\ref{masssplit}).
They form the  fundamental representation of the global SU$(N)$ group in this limit.

Much in the same way as in the $\Lambda$-vacua, the total number 
of the  kinks interpolating between each pair of the neighboring zero-vacua
is $N_f$.
They form the $(N,1)+(1,\tN)$  representation of the global group (\ref{c+fd}) in the 
limit (\ref{masssplit}). Note, that the $P$-kinks in the zero-vacua are heavier than the $K$-kinks.
Their masses are given by $\Delta m$ (to the leading order) while the $K$-kink masses 
are of the order of $\tilde{\Lambda}_{LE}$. Still, given
small hierarchical masses (\ref{smallmasses}),
all kinks in the zero-vacua are lighter 
than
those  in the $\Lambda$-vacua which have masses of the order of $\Lambda$, see Eqs.~(\ref{LspectrP})
and (\ref{LspectrK}).

In parallel with the $\Lambda$-vacua, we can verify that Eq.~({\ref{BPSmass}) reproduces
the above BPS spectrum with the appropriate choice of the integration contours. Namely,
for the $P_0$-kink the contour in the $\sigma$ plane encircles the pole at 
$\sqrt{2}\sigma=-m_{P_0}$ in the anticlockwise direction, cf. Fig.~\ref{figcontourcp}.
 For the $K_0$-kink
the contour encircles the pole at 
$\sqrt{2}\sigma=-m_{K_0}$ in the clockwise direction.

\section{Lessons for the bulk theory}
\label{secmoral}
\setcounter{equation}{0}

This section carries a special weight and is, in a sense, central for the present investigation,
since here
 we translate our results for the kink spectrum in the weighted two-dimensional CP model
(\ref{wcp}) in the language of strings and confined monopoles of the bulk four-dimensional theory.

We start from the most interesting strong-coupling domain 
\beq
\xi\ll \Lambda
\label{strongcoupling}
\eeq
which can be  described in terms of weakly-coupled dual bulk theory \cite{SYdual},
see Sec.~\ref{secbulkduality}. At this point we take the limit (\ref{masssplit}) to ensure
the presence of the unbroken global group (\ref{c+fd}).

As was mentioned previously, the elementary non-Abelian strings of the bulk theory correspond
to various vacua of the world-sheet two-dimensional theory, see, e.g. the review paper~\cite{SYrev}
for a detailed discussion. The weighted CP model  (\ref{dcp}) has two types of 
vacua, namely: $(N-\tN)$ $\Lambda$-vacua and $\tN$ zero-vacua. The former   are not-so-interesting
from the standpoint of the bulk theory. Indeed, they yield just Abelian $Z_{N-\tN}$ strings associated with 
the winding of the $(N-\tN)$ singlet dyons $D^l$ charged with respect to U(1) 
factors of the gauge group of 
the dual theory (\ref{dualgaugegroup}) \cite{SYdual}, see (\ref{Dvev}). 
Moreover,
the weighted CP model  (\ref{dcp}) is, in fact, inapplicable in the description of these strings.
This model is an effective low-energy theory which can be used below the scale $\sqrt{\xi}$.
However, the energy scale in the $\Lambda$-vacua is of order of $\Lambda$, i.e.  much
larger than $\sqrt{\xi}$ in the domain (\ref{strongcoupling}).

Below we focus on $\tN$ zero-vacua which correspond to $\tN$ elementary {\em non}-Abelian
strings  associated with the winding of the light dyons $D^{lA}$ of the dual bulk theory.
The latter are
charged with respect to both Abelian and non-Abelian factors  \cite{SYdual} of the dual 
gauge group (\ref{dualgaugegroup}). The energy scale in these vacua
of the world-sheet theory is of the order of 
$$
{\rm max}(\Delta m_{KK'},\tilde{\Lambda}_{LE})\,,
$$
which we assume to be much less than $\sqrt{\xi}$. Thus, in these vacua the weighted CP model (\ref{dcp})
can be applied   to describe the internal dynamics of the non-Abelian strings of the bulk theory.

The confined monopoles of the bulk theory are
seen as kinks in the world-sheet theory. 
The results presented in Sec.~\ref{seckink}  demonstrate that in 
the weighted CP model
there are $N_f$ elementary kinks  interpolating between 
the neighboring zero-vacua. More exactly, we found  $N$ $P$-kinks with masses (\ref{ZspectrP}) and 
$\tN$ $K$-kinks with masses (\ref{ZspectrK}). In the limit (\ref{masssplit}) they
form the $(N,1)+(1,\tN)$  representations of the global group (\ref{c+fd}).

This means that the total number of stringy mesons $M_A^B$ formed by 
the monopole-antimonopole
pairs connected by two different elementary non-Abelian strings  (Fig.~\ref{figmeson}) is  $N_f^2$.
The mesons $M_P^{P'}$ form the
singlet and $(N^2-1,1)$ adjoint representations  of the global group (\ref{c+fd}), 
the mesons $M_P^{K}$ and $M_K^{P}$ form bifundamental representations
$(N,\bar{\tN})$ and $(\bar{N},\tN)$, while the mesons 
$M_K^{K'}$ form the singlet and $(1,\tN^2-1)$ adjoint representations. 
(Here as usual, $P=1,...,N$ and $K=(N+1),...,N_f$.)
All these mesons have masses of the order of $\sqrt{\xi}$, determined by the 
string tension
\beq
T=2\pi\xi\,.
\label{ten}
\eeq
 They are heavier than the
elementary states, namely, dyons and dual gauge bosons which form 
the (1,1), $(N,\bar{\tN})$, $(\bar{N},\tN)$, 
and $(1,\tN^2-1)$ representations and have masses $\sim \tilde{g}_2\sqrt{\xi}$.

Therefore, the  (1,1), $(N,\bar{\tN})$, $(\bar{N},\tN)$, and $(1,\tN^2-1)$ stringy mesons
decay into elementary states, and we are left with  $M_P^{P'}$ stringy mesons
in the representation 
$(N^2-1,1)$.
This is exactly what was predicted in \cite{SYdual} from the bulk perspective, see
Sec.~\ref{secbulkduality}. Thus, our world sheet picture nicely matches the bulk analysis.

Note also that the $M_P^{P'}$ stringy mesons with strings   corresponding to the
$\Lambda$-vacua of the weighted CP model (the ``$\Lambda$-strings") are heavy and decay into 
$M_P^{P'}$ stringy mesons with strings   corresponding to
the zero-vacua (the ``zero-strings"). To see that this is indeed the case, observe that 
the confined monopoles (i.e. kinks
of the weighted CP model) on the $\Lambda$-string have masses of the order of $\Lambda$,
see (\ref{LspectrP}). Therefore, the $\Lambda$-stringy  mesons  also have
masses of the order of $\Lambda$. The  $M_P^{P'}$  mesons with the zero-strings
  are much lighter in the domain (\ref{strongcoupling}). Their masses are of the order of 
${\rm max}(\Delta m,\sqrt{\xi})$.

Now, let us discuss yet  another match of the world-sheet and bulk pictures.  
In Sec.~\ref{seclargemasses} we saw that there are elementary excitations 
($T=0$ and $S_A=\delta_{AP_1}-\delta_{AP_2}$)   on the
string at weak coupling (this is attainable with   large
$|\Delta m_{AB}|$). These excitations would form the  adjoint representation
$(N^2-1,1)$ of the global group if the limit (\ref{masssplit}) could
be taken. However, in the strong coupling domain of hierarchically
small masses (\ref{smallmasses}) the kink spectrum is very different, see 
Sec.~\ref{seckink}. In particular, no elementary excitations are left: these 
states decay
on CMS into a $P_1$-kink plus a $P_2$-antikink, see Eqs.~(\ref{LspectrP}) and  (\ref{ZspectrP}). 

Since the BPS
spectra of the \ntwot two-dimensional theory on the string and 
the \ntwo four-dimensional
bulk theory on the Coulomb branch (at zero $\xi$) coincide \cite{Dorey,DoHoTo,SYmon,HT2},
the decay process above is in one-to-one correspondence with the decay of the bulk states
identified in \cite{SYdual}.
Namely, 
the quarks $q^{kP_1}$ (with $k=P_2$ due to 
the color-flavor locking) and the gauge bosons present in the bulk theory at weak coupling
decay into monopole-antimonopole pairs. If $\xi$ is small but nonvanishing, the monopoles
and antimonopoles cannot move apart: they are bound together by pairs of 
confining strings and form  \cite{SYdual}
the $M_{P_2}^{P_1}$ mesons  shown in Fig.~\ref{figmeson}. Thus, our results from two dimensions confirm the
decay of the quarks and gauge bosons in the strong-coupling domain of the bulk theory.
 This decay
process is a crucial element of our mechanism of non-Abelian confinement.

To explain this in more detail we present a ``phase diagram" of the bulk
theory,  see Fig.~\ref{figphdiag}. The vertical and horizontal solid
lines in this figure schematically represent the bulk theory
CMS. The horizontal axis gives the masses 
$\Delta m_{PP'}\sim \Delta m_{KK'}$ which we force to be of the same order,
while $\Delta m$ is fixed, $\Delta m \ll \Lambda$. The vertical axis gives
the FI parameter $\xi$. The vertical dashed lines depict CMS of the world-sheet theory.
On these lines the spectrum of the stringy mesons of the bulk theory rearranges itself.
 On the solid lines the ``perturbative"
spectrum of the bulk theory rearranges itself. Different domains inside CMS
(where the spectra change  continuously) are denoted by capital letters  A, B, ..., F.

\begin{figure}
\epsfxsize=10cm
\centerline{\epsfbox{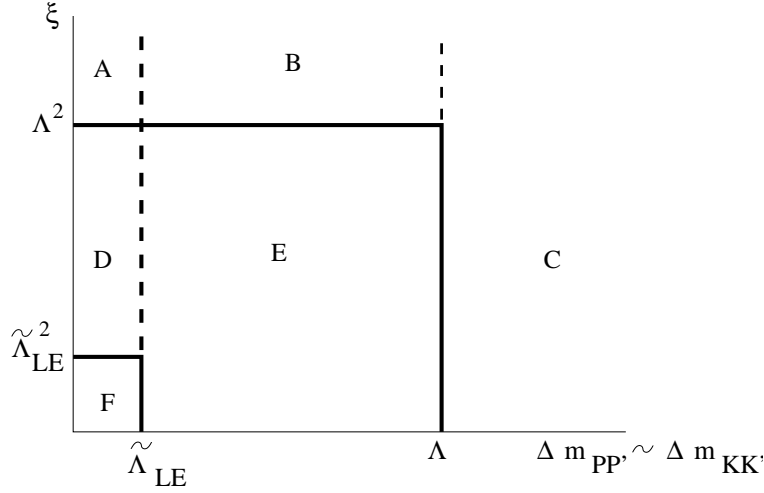}}
\caption{\small ``Phase diagram" of the bulk theory. Various domains
are separated by CMS on which physical spectrum is rearranged.}
\label{figphdiag}
\end{figure}

The domains $A$, $B$ and $C$ are at weak coupling in the original U$(N)$ gauge
theory. The elementary (``perturbative")  states in these domains are the $q^{kA}$
quarks  and gauge bosons
of the U($N$) gauge group. The domains $D$ and $E$ are at weak coupling  in the dual theory,
see Sec.~\ref{secbulkduality}. The elementary (``perturbative") states in these domains  
are the light dyons
$D^{lA}$ and  $D^l$ in Eq.~(\ref{Dvev}) plus the gauge bosons of the dual gauge group (\ref{dualgaugegroup}).
Masses of all these states smoothly evolve with the change of parameters inside these
domains. The domain $F$ is at strong coupling   in the dual theory. 
 With small mass differences in the domain
(\ref{smallmasses}), $N$ flavors of dyons in Eq.~(\ref{Dvev}) decouple, the dual theory becomes
asymptotically free, and at  $\xi\ll \tilde{\Lambda}_{LE}$ passes into the strong coupling regime.
The dual gauge group (\ref{dualgaugegroup}) gets broken down to U(1)$^N$ by the 
Seiberg--Witten mechanism.

Keeping in mind the limit (\ref{masssplit}), our task is to pass from the domain $A$ into the domain
$D$ and prove that the quarks and gauge bosons of the former domain decay into 
the monopole-antimonopole
pairs connected by confining strings in the domain $D$. We do it in three steps. 
First,
we pass from the domain $A$ to  $C$ at large $\xi$
and then move towards small $\xi$ (preserving large mass differences) inside $C$.
The original bulk theory is in the weak-coupling regime in these domains and the quarks and gauge boson spectra
evolve smoothly.

Next,
 we pass from the domain $C$ to the domain $E$ at small (or vanishing) $\xi$. Here we 
use correspondence between the
BPS spectra of the bulk and world-sheet theories. The
 $q^{kP_1}$ quarks (with $k=P_2$ due to the
 color-flavor locking) and gauge bosons of the domain $C$ correspond to elementary states 
with $T=0$ and $S_A=\delta_{AP_1}-\delta_{AP_2}$, see (\ref{elmass}). As was already
explained, these states decay on CMS of the two-dimensional theory into a
$P_1$-kink and a $P_2$-antikink, interpolating between the $\Lambda$-vacua, see (\ref{LspectrP}). 
Since the spectra
of the massive BPS  states in the bulk (at $\xi=0$) on the one hand, and in the world-sheet theory
on the other hand,   are identical, both the
quarks and gauge bosons from domain $C$ do not  exist in the domain $E$. They decay
into a $P_1$-monopole and a $P_2$-antimonopole. 
At small but  nonvansishing $\xi$ the latter
states are confined by $\Lambda$-strings which gives rise to  $M_{P_2}^{P_1}$ stringy mesons.

Furthermore, as  we pass from the domain $E$ to our final destination --  the domain $D$ -- these $M_{P_2}^{P_1}$
mesons ``glued" by $\Lambda$-strings,  decay on   the world-sheet theory CMS   located 
at $\Delta m_{PP'}\sim \Delta m_{KK'}\sim \tilde{\Lambda}_{LE}$.
They decay  into lighter $M_{P_2}^{P_1}$ mesons glued by zero-strings. The latter stringy mesons were 
absent  in the domain
$E$ at intermediate masses (see (\ref{kinkspectrim})), but emerge in the domain $D$
at small masses, see (\ref{ZspectrP}). This completes our proof.

In conclusion it is worth noting that
the stringy mesons $M_P^{P'}$ in the adjoint representation  $(N^2-1,1)$
of the global group
are metastable, strictly speaking. An extra monopole-antimonopole pair can be
created on the string, making the $(N^2-1,1)$ meson to decay into a pair of 
stringy mesons in the bifundamental representation,  $(N,\bar{\tN})$ and $(\bar{N},\tN)$.
During the subsequent stage these stringy bifundamentals decay into elementary
bifundamental dyons. One  can suppress the rate of this decay, however. If we
keep $\Delta m\ll \sqrt{\xi}$ and take a limit of large $N$, while $\tN$ is fixed
this decay rate is of the order of $\tN/N\ll 1$. To see that this indeed the case, note that the above
decay process goes through  creation of a monopole-antimonopole pair  from
the fundamental representation of SU($\tN$) on the string which selects
 $\tN$ channels out of $N_f$.

Similar considerations can be applied to the weak-coupling domain of large
$\xi$, 
$$\sqrt{\xi}\gg \Lambda\,,
$$ of the bulk theory. We still have kinks
in the $(N,1)+(1,\tN)$  representations of the global group (\ref{c+f})
in the limit (\ref{masssplit}). Thus, all types of mesons $M^A_B$ are formed
with masses $\sim\sqrt{\xi}$.
However, in this regime the
(1,1), $(N,\bar{\tN})$, $(\bar{N},\tN)$, and $(N^2-1,1)$ stringy mesons
can decay into quarks and gauge bosons 
with the same quantum numbers and masses $\sim g_2^2\sqrt{\xi}$. Therefore,
 we are left with elementary states and stringy adjoint mesons $(1,\tN^2-1)$. 
The latter  are metastable and decay in pairs 
of bifundamental states.

\section{Conclusions}
\label{concl}

Both \ntwo four-dimensional theories belonging to the dual pair discussed above
support non-Abelian strings. The world-sheet theory on the strings is given by the weighted CP
model which appears in two varieties depending on which side of duality we are. These two
weighted CP
models also form a dual pair.  We explore the kink spectra 
(which represent confined monopoles of the bulk theory) and their evolution
in passing through the crossover domain. Of most interest is
small-$\xi$ dynamics. In fact, at small $\xi$ we find {\em two} weak coupling subdomains and 
a strong-coupling one depending on the values of the differences of the mass parameters
$\Delta m_{AB}$. 

We have shown that in the limit (\ref{masssplit}) where the global group
(\ref{c+f}) is unbroken confined monopoles form the fundamental representation
of the global group.  Therefore, stringy mesons
(shown in Fig.~\ref{figmeson}) formed by pairs of monopoles and antimonopoles
belong to the adjoint or singlet representations of the global group. This nicely matches
global quantum numbers of mesons in the ``real world."

We proved the statement proposed in \cite{SYdual} that quarks and gauge
bosons present in the original theory at large $\xi$ decay on CMS into 
monopole-antimonopole pairs confined by non-Abelian strings as we enter the small-$\xi$ domain.
This result is a crucial element of our mechanism of non-Abelian confinement.

In summary, in this paper we used the world-sheet theory to confirm the picture of non-Abelian
 confinement obtained in \cite{SYdual} from the bulk perspective. Non-Abelian confinement 
is {\em not} associated with formation of chromoelectric strings connecting quarks, as 
a naive extrapolation of the Abelian confinement picture suggests. 
Rather, it is due to the decay on CMS of the Higgs-screened quarks and 
gauge bosons into monopole-antimonopole pairs confined by non-Abelian strings
in the strong coupling domain of small $\xi$. We stress again that the non-Abelian strings confine
 monopoles
both in the original and dual theories.\footnote{ Similar results were recently obtained
in \cite{MarsY} for \ntwo supersymmetric QCD with $N_f>2N$.}

Analysis of the mass spectra presented in the bulk of the paper raises a number of intriguing questions. 
One of them refers to typical sizes of the objects considered. At the moment, in the absence of a detailed analysis,
one can address this issue only at the qualitative level. At small $\xi$ and $\Delta m 
\ll \sqrt{\xi}\ll \Lambda$
we expect that
the smallest size $\sim \Lambda^{-1}$ is that of the elementary (``perturbative") states,
namely dyons from Eq.~(\ref{Dvev}) and dual gauge bosons.
The sizes of the stringy mesons in the representation $N^2-1$ 
are expected to be of the order $(\tilde g \sqrt{\xi})^{-1}$, i.e of order of  the thickness of the non-Abelian strings.
Finally, the largest sizes $(\Delta m)^{-1}$ and  $\sim \tilde\Lambda_{LE}^{-1}$
belongs to the $P$- and $K$-kinks (confined monopoles on a string), respectively,  provided that the mass difference
$\Delta m_{KK^\prime} \to 0$.

\section*{Acknowledgments}
We are grateful to A. Gorsky for valuable discussions.

This work  is supported in part by DOE grant DE-FG02-94ER408. 
The work of A.Y. was  supported 
by  FTPI, University of Minnesota, 
by RFBR Grant No. 09-02-00457a 
and by Russian State Grant for 
Scientific Schools RSGSS-11242003.2.

\vspace{1cm}


\section*{Appendix A: \\
Kink solutions in \boldmath{\cpn} model}

 \renewcommand{\theequation}{A.\arabic{equation}}
\setcounter{equation}{0}
 
 \renewcommand{\thesubsection}{A.\arabic{subsection}}
\setcounter{subsection}{0}

In this appendix we discuss in more detail the kink solutions in 
the $\mbox{\cpn}$ model (see Sec.~\ref{seccpnkinks}).
For  simplicity we set  $\Delta m_{PP'}=0$. Small mass differences  will 
slightly  deform the kink profile functions    but will not change the very
fact of its existence. The kink masses   are given by differences
of the mirror superpotential evaluated at the initial and final vacua, see (\ref{cpspectr}).

We  look for the kink solution interpolating between $l=0$ and $l=1$ vacua
 using the {\em ansatz}
\beq
X_{P\neq P_0}=r\,e^{i\theta},\qquad X_{P_0}=\frac{e^{-i(N-1)\theta}}{r^{(N-1)}}\,,
\label{cpansatz}
\eeq
where $r(z)$ and $\theta(z)$ are kink profile functions, subject to boundary
conditions
\beqn
\theta(z=-\infty)
&=&
0,\qquad r(z=-\infty)=1,
\nonumber\\[3mm]
\theta(z=\infty)
&=&
\frac{2\pi}{N},\qquad r(z=\infty)=1,
\label{cpbc}
\eeqn
see (\ref{xvevcp}). The last expression in (\ref{cpansatz}) is dictated by
(\ref{constraintcp}).

The explicit form of the kink profile functions depends on the form of the kinetic term
which is not known. Therefore, both profile functions $r(z)$ and $\theta(z)$
cannot be determined. Still we can obtain the kink solution up to a single unknown
profile function $\theta(z)$ expressing $r$ as a function of $\theta$.
To this end we exploit the fact that the kink trajectory in the complex plane of  superpotential
goes along the straight line connecting the points $W_{\rm mirror}(l=0)$ with
$W_{\rm mirror}(l=1)$ \cite{ChVa}.  The difference 
\beqn
&&
W_{\rm mirror}(l=1)-W_{\rm mirror}(l=0)=
-\frac{N\Lambda}{4\pi}\left(e^{\frac{2\pi}{N}\,i}
-1\right)
\nonumber\\[3mm]
&&
=-\frac{N\Lambda}{4\pi}\,e^{\frac{\pi}{N}\,i}\,2i\sin{\left(\frac{\pi}{N}\right)},
\label{cpdiff}
\eeqn
where we used (\ref{mirrorcp}) and (\ref{xvevcp}).

\vspace{2mm}

On the other hand,
\beqn
&&
W_{\rm mirror}(\theta)-W_{\rm mirror}(\theta=0)=
-\frac{\Lambda}{4\pi}\left[ (N-1)re^{i\theta}+\frac1{r^{N-1}}e^{-i(N-1)\theta}-N\right]
\nonumber\\[3mm]
&&
=-\frac{r\Lambda}{4\pi}\,e^{\frac{\pi}{N}\,i}\left\{(N-1)e^{i(\theta-\frac{\pi}{N})}
+\frac1{r^{N}}e^{-i(N-1)\theta-i\frac{\pi}{N}}-\frac{N}{r}e^{-\frac{\pi}{N}}\right\}.
\label{cpdiffp}
\eeqn
Comparing Eqs.~(\ref{cpdiff}) and (\ref{cpdiffp}) we see that  for the kink 
trajectory to go along the straight line, the real part of the expression in the curly brackets
in (\ref{cpdiffp}) must vanish. As a result,
\beq
(N-1)\cos{\left(\theta-\frac{\pi}{N}\right)}-\frac{N}{r}\cos{\frac{\pi}{N}}
+\frac1{r^{N}}\cos{\left[(N-1)\theta+\frac{\pi}{N}\right]}=0.
\label{cpkinkeq}
\eeq
This equation determines the function $r(\theta)$. It   has
a nonvansihing nonsingular solution at $0\leq \theta\leq 2\pi/N$ which satisfies
the boundary conditions $r(2\pi/N)=r(0)=1$. Say, for $N=2$,  $r=1$.  For large $N$
the profile function $r$ is approximately given by
\beq
r(\theta)\approx 1+\frac1{N}\left\{1-\cos{\left[(N-1)\theta+\frac{\pi}{N}\right]}
\right\}+\cdots .
\label{largeNrcp}
\eeq

\section*{Appendix B: \\
Kink solutions in the zero-vacua}

 \renewcommand{\theequation}{B.\arabic{equation}}
\setcounter{equation}{0}

Here we consider the
kink solutions interpolating between the zero-vacua with $l=0$ and $l=1$,
see (\ref{Yvev}), at hierarchically small masses (\ref{smallmasses}). The
$K$-kink solutions in the zero-vacua are quite similar to kinks in the \cpn model.
Therefore, we focus on $P$-kinks.

\begin{figure}
\epsfxsize=8cm
\centerline{\epsfbox{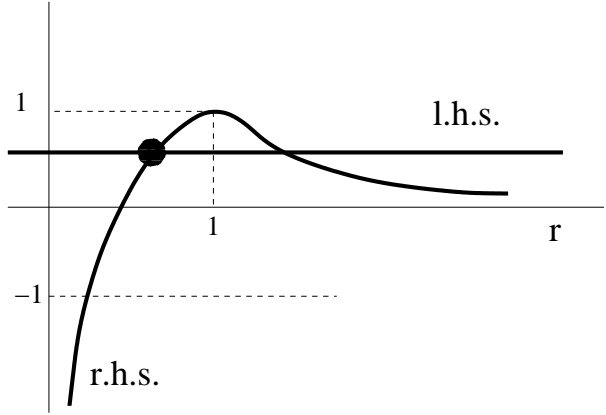}}
\caption{\small The right- and left-hand sides of Eq.~(\ref{kinkeq}). The closed circle denotes the regular
solution. }
\label{figkinkeq}
\end{figure}

Analyzing the $P$-kinks we
assume the limit (\ref{masssplit}) (for simplicity). The {\em ansatz} for the kink
solution takes the form
\beqn
Y_{K}
&=&
r\,e^{i\theta}\,\,\frac{\tilde{\Lambda}_{LE}}{\Lambda}\,,
\nonumber\\[3mm]
X_{P\neq P_0}
&\approx&
 \frac{\Delta m}{\Lambda},\qquad \quad \,\,\, 
 X_{P_0}\approx\frac{\Delta m}{\Lambda}\,r^{\tN}e^{i\tN \theta}\,,
\label{ansatz}
\eeqn
where we restrict ourselves to the leading order contributions ($\sim \Delta m/\Lambda$)
ignoring next-to-leading $O(\tilde{\Lambda}_{LE}/\Lambda )$ terms. The profile functions 
$r(z)$ and $\theta(z)$ are subject to the  boundary conditions 
\beqn
\theta(z=-\infty)
&=&
0,\qquad r(z=-\infty)=1\,,
\nonumber\\[3mm]
\theta(z=\infty)
&=&
\frac{2\pi}{\tN},\qquad r(z=\infty)=1\,,
\label{bc}
\eeqn
see (\ref{Yvev}) and (\ref{Xvevzero}). The last expression in (\ref{ansatz}) is dictated by
(\ref{constraint}).

The superpotential
(\ref{mirror}) then gives
\beqn
&&
W_{\rm mirror}(\theta)-W_{\rm mirror}(\theta=0)
\nonumber\\[3mm]
&&
\approx-\frac{\Delta m}{4\pi}\left\{r^{\tN}e^{i\tN \theta}
-\tN\,(i\theta +\ln{r})-1\right\}\,,
\label{supoftheta}
\eeqn
while the difference of the superpotential in the initial and final vacua is
\beq
W_{\rm mirror}(l=1)-W_{\rm mirror}(l=0)\approx
\frac{\Delta m}{4\pi}\times 2\pi i\,.
\label{supofthetap}
\eeq
For the kink trajectory to go along the straight line in the complex plane  of
superpotential,  the real part of the expression in the curly brackets
in (\ref{supoftheta}) must vanish. This requirement implies
\beq
r^{\tN}\cos{(\tN \theta)}-\tN\ln{r}=1\,.
\label{kinkeq}
\eeq
The latter  equation always has a finite nonvanishing solution
in the interval  $0\leq \theta\leq 2\pi/N$, subject to the
boundary conditions $r(2\pi/N)=r(0)=1$. To 
check that this is indeed the case
we rewrite it as
\beq
\cos{(\tN \theta)}=\frac{1+\tN\ln{r}}{r^{\tN}}\,.
\label{dop9}
\eeq
The right- and left-hand sides of (\ref{dop9})  are schematically plotted
in Fig.~\ref{figkinkeq}. For any $-1<\cos{(\tN \theta)}<1$ we have only one 
nonsingular solution $r(\theta)$. In particular, at $\tN\gg 1$
\beq
r(\theta)\approx 1-\frac1{\tN}[1-\cos{(\tN \theta)}]+\cdots\,.
\label{dop9p}
\eeq


\newpage 
\small
\begin{thebibliography}{99}
\addcontentsline{toc}{section}{References}
\itemsep -2pt


\bibitem{mandelstam}
Y.~Nambu,
  Phys.\ Rev.\  D {\bf 10}, 4262 (1974);\\
G.~'t Hooft,
{\em Gauge theories with unified weak, electromagnetic and strong interactions,}
in Proc. of the E.P.S. Int. Conf. on High Energy Physics, Palermo, 23-28 June, 1975
ed. A. Zichichi (Editrice Compositori, Bologna, 1976);
Nucl.\ Phys.\ B {\bf 190}, 455 (1981);
S.~Mandelstam,
Phys.\ Rept.\  {\bf 23}, 245 (1976).


\bibitem{ANO}
A.~Abrikosov, Sov.~Phys. JETP {\bf32}, 1442  (1957)
[Reprinted in {\em Solitons and Particles}, Eds. C. Rebbi and G. Soliani
(World Scientific, Singapore, 1984), p. 356];
H.~Nielsen and P.~Olesen, Nucl.~Phys. {\bf B61}, 45 (1973)
[Reprinted in {\em Solitons and Particles}, Eds. C. Rebbi and G. Soliani
(World Scientific, Singapore, 1984), p. 365].

\bibitem{SW1}
N.~Seiberg and E.~Witten,
Nucl. Phys. {\bf B426}, 19 (1994),
(E) {\bf B430},  485 (1994) [hep-th/9407087].

 \bibitem{SW2}
N.~Seiberg and E.~Witten,
Nucl. Phys. {\bf B431}, 484  (1994)
[hep-th/9408099].

\bibitem{Polyakov}
  A.~M.~Polyakov,
  Nucl.\ Phys.\  B {\bf 120}, 429 (1977).

\bibitem{DS}
M.~R.~Douglas and S.~H.~Shenker,
Nucl.\ Phys.\ B {\bf 447}, 271 (1995)
[hep-th/9503163].

\bibitem{HSZ}
A.~Hanany, M.~J.~Strassler and A.~Zaffaroni,
Nucl.\ Phys.\ B {\bf 513}, 87 (1998)
[hep-th/9707244].

\bibitem{Strassler}
M.~Strassler,
  Prog.\ Theor.\ Phys.\ Suppl.\  {\bf 131}, 439 (1998)
  [hep-lat/9803009].

\bibitem{VY}
A.~I.~Vainshtein and A.~Yung,
Nucl.\ Phys.\ B {\bf 614}, 3 (2001)
[hep-th/0012250].

\bibitem{Yrev}
A.~Yung,
{\em What Do We Learn About Confinement From The Seiberg--Witten Theory?},
Proc. of 28th PNPI Winter School of Physics, St. Petersburg, Russia, 2000,
[hep-th/0005088];  published in  {\sl At the frontier of particle physics}, Ed. M.~Shifman,
(World Scientific, Singapore, 2001)
vol. 3, p. 1827.

   \bibitem{SYdual}
M.~Shifman and A.~Yung, 
Phys. \ Rev. \ D {\bf 79}, 125012 (2009)
[arXiv:0904.1035 [hep-th]].

   \bibitem{FI}
  P.~Fayet and J.~Iliopoulos,
  Phys.\ Lett.\  B {\bf 51}, 461 (1974).

  \bibitem{HT1}
A.~Hanany and D.~Tong,
JHEP {\bf 0307}, 037 (2003)
[hep-th/0306150].

\bibitem{ABEKY}
R.~Auzzi, S.~Bolognesi, J.~Evslin, K.~Konishi and A.~Yung,
Nucl.\ Phys.\ B {\bf 673}, 187 (2003)
[hep-th/0307287].

 \bibitem{SYmon}
M.~Shifman and A.~Yung,
Phys.\ Rev.\ D {\bf 70}, 045004 (2004)
[hep-th/0403149].

\bibitem{HT2}
A.~Hanany and D.~Tong,
JHEP {\bf 0404}, 066 (2004)
[hep-th/0403158].
  
 \bibitem{Trev}
D.~Tong, {\em TASI Lectures on Solitons,}
  arXiv:hep-th/0509216.

\bibitem{Jrev}
  M.~Eto, Y.~Isozumi, M.~Nitta, K.~Ohashi and N.~Sakai,
  J.\ Phys.\ A  {\bf 39}, R315 (2006)
  [arXiv:hep-th/0602170].
  
  \bibitem{SYrev}
M.~Shifman and A.~Yung,
{\sl Supersymmetric Solitons,}
Rev.\ Mod.\ Phys. {\bf 79} 1139 (2007)
[arXiv:hep-th/0703267]; an expanded version in Cambridge University Press, 2009.

\bibitem{Trev2}
D.~Tong,
  Annals Phys.\  {\bf 324}, 30 (2009)
  [arXiv:0809.5060 [hep-th]].

  \bibitem{SYcross}
M.~Shifman and A.~Yung,
Phys. Rev. {\bf D 79}, 105006 (2009)
  arXiv:0901.4144 [hep-th].

\bibitem{SYcrossp}
M.~Shifman and A.~Yung,
{\em Non-Abelian Strings: From Weak to Strong Coupling and Back via Duality,}
  arXiv:0910.3007 [hep-th].
  
  \bibitem{APS}
P.~Argyres, M.~Plesser and N.~Seiberg,
Nucl. Phys. {\bf B471}, 159  (1996)
[hep-th/9603042].
 
\bibitem{Sdual}
  N.~Seiberg,
  Nucl.\ Phys.\  B {\bf 435}, 129 (1995)
  [arXiv:hep-th/9411149].
    
\bibitem{IS}
K.~A.~Intriligator and N.~Seiberg,
  Nucl.\ Phys.\ Proc.\ Suppl.\  {\bf 45BC}, 1 (1996)
  [hep-th/9509066].

\bibitem{Dorey}
N.~Dorey,
JHEP {\bf 9811}, 005 (1998) [hep-th/9806056].

\bibitem{DoHoTo}
N.~Dorey, T.~J.~Hollowood and D.~Tong,
  JHEP {\bf 9905}, 006 (1999)
  [arXiv:hep-th/9902134].
  
\bibitem{AchVas}
A.~Achucarro and T.~Vachaspati,
  Phys.\ Rept.\  {\bf 327}, 347 (2000)
  [hep-ph/9904229].

\bibitem{SYsem}
 M.~Shifman and A.~Yung,
  Phys.\ Rev.\  D {\bf 73}, 125012 (2006)
  [arXiv:hep-th/0603134].

\bibitem{Jsem}
M.~Eto, J.~Evslin, K.~Konishi, G.~Marmorini, M.~Nitta, K.~Ohashi, W.~Vinci, N.~Yokoi,
  Phys.\ Rev.\  D {\bf 76}, 105002 (2007)
  [arXiv:0704.2218 [hep-th]].
  
\bibitem{W79}
E.~Witten,
Nucl.\ Phys.\ B {\bf 149}, 285 (1979).

 \bibitem{HoVa}
  K.~Hori and C.~Vafa,
{\em Mirror symmetry,}
  arXiv:hep-th/0002222.
  
\bibitem{BF}
A.~Bilal and F.~Ferrari,
  Nucl.\ Phys.\  B {\bf 516}, 175 (1998)
  [arXiv:hep-th/9706145].
  
 \bibitem{T}
D.~Tong,
Phys.\ Rev.\ D {\bf 69}, 065003 (2004)
[hep-th/0307302].

\bibitem{W93}
E.~Witten,
  Nucl.\ Phys.\ B {\bf 403}, 159 (1993)
  [hep-th/9301042].
  
\bibitem{EY}
 K.~Evlampiev, A.~Yung,
 Nucl.\ Phys. \ B {\bf 662}, 120  (2003)
 [hep-th/0303047]. 
   
\bibitem{AdDVecSal}
A.~D'Adda, A.~C.~Davis, P.~DiVeccia and P.~Salamonson,
Nucl.\ Phys.\ {\bf B222} 45 (1983).
   
\bibitem{ChVa}
S.~Cecotti and C. Vafa,
Comm. \ Math. \ Phys. \ {\bf 158} 569 (1993).

\bibitem{VYan}
 G.~Veneziano and S.~Yankielowicz,
  Phys.\ Lett.\  B {\bf 113}, 231 (1982).

\bibitem{HaHo}
A.~Hanany and K.~Hori,
  Nucl.\ Phys.\  B {\bf 513}, 119 (1998)
  [arXiv:hep-th/9707192].
[hep-th/9211097].

  \bibitem{svz}
M.~Shifman, A.~Vainshtein and R.~Zwicky,
  J.\ Phys.\ A  {\bf 39}, 13005 (2006)
  [arXiv:hep-th/0602004].
  
\bibitem{OlmezS}
S.~\"Olmez and M.~Shifman,
J. \ Phys. \ A {\bf 40}, 11151 (2007)
[hep-th/0703149]

\bibitem{LeeYi}
S.~Lee, P.~Yi, {\em A Study of Wall-Crossing: Flavored Kinks in D=2 QED},
arXiv:0911.4726 [hep-th].
  
\bibitem{FFS}
V.~A.~Fateev, I.~V.~Frolov and A.~S.~Schwarz,
Sov.\ J.\ Nucl.\ Phys.  {\bf 30}, 590 (1979)
[Yad.\ Fiz.  {\bf 30}, 1134 (1979)];
Nucl.\ Phys.\ B {\bf 154} (1979) 1.
See also in A. Polyakov, {\sl Gauge Fields and Strings}
(Harwood Press, 1987).

\bibitem{RitzSV}
A.~Ritz, M. Shifman and  A. Vainshtein,
Phys. \ Rev. \ D {\bf 66}, 065015 (2002)
[hep-th/0205083].

\bibitem{MarsY}
A.~Marshakov and A.~Yung,
{\em Strong versus weak coupling confinement in \ntwo supersymmetric QCD},
arXiv:0912.1366 [hep-th].


\end{thebibliography}
\end{document}